\documentclass[a4paper,showpacs,12pt]{article}

\usepackage{slashed}

\usepackage{amsmath}\usepackage{amssymb}
\usepackage{latexsym}\usepackage{cite}

\usepackage{amsmath}
\usepackage[dvips]{graphicx}
\usepackage{pstricks}
\usepackage{color}
\usepackage{bm}

\usepackage{graphicx}

\definecolor{Red}{rgb}{1.00, 0.00, 0.00}
\definecolor{Green}{rgb}{0.00, 1.00, 0.00}
\definecolor{Blue}{rgb}{0.00, 0.00, 1.00}
\definecolor{Cyan}{rgb}{0.00, 1.00, 1.00}
\definecolor{Mymagenta}{rgb}{0.3, 0.00, 1.00}%
\definecolor{Magenta}{rgb}{1.00, 0.00, 1.00}
\definecolor{DeepSkyBlue}{rgb}{0.00, 0.75, 1.00}
\definecolor{DarkGreen}{rgb}{0.00, 0.39, 0.00}
\definecolor{SpringGreen}{rgb}{0.00, 1.00, 0.50}
\definecolor{Mygreen}{rgb}{0.00, 0.72, 0.00}
\definecolor{DarkOrange}{rgb}{1.00, 0.55, 0.00}
\definecolor{OrangeRed}{rgb}{1.00, 0.27, 0.00}
\definecolor{DeepPink}{rgb}{1.00, 0.08, 0.57}
\definecolor{DarkViolet}{rgb}{0.58, 0.00, 0.82}
\definecolor{SaddleBrown}{rgb}{0.57, 0.27, 0.07}
\definecolor{Black}{rgb}{1.00, 1.00, 1.00}
\definecolor{Ablue}{rgb}{0.10, 0.1, 1.00}

\newcommand{\be}{\begin{equation}}
\newcommand{\ee}{\end{equation}}

\newcommand{\ka}{\kappa}

\def\beq{\begin{equation}}
\def\eeq{\end{equation}}
\def\beqr{\begin{eqnarray}}
\def\eeqr{\end{eqnarray}}
\def\pl{\partial}

\def\al{\alpha}
\def\bt{\beta}
\def\Ga{\Gamma}
\def\ga{\gamma}
\def\de{\delta}
\def\De{\Delta}

\def\ka{\kappa}

\def\Si{\Sigma}
\def\te{\theta}

\def\La{\Lambda}
\def\lam{\lambda}
\def\Om{\Omega}
\def\om{\omega}

\def\sq{\sqrt}

\def\l{\left (}
\def\r{\right )}

\def\fr{\frac}
\def\la{\label}
\def\hs{\hspace}
\def\vs{\vspace}

\def\ran{\rangle}
\def\lan{\langle}
\def\ov{\overline}
\def\tl{\tilde}
\def\tm{\times}

\def\lrarr{\longrightarrow}

\begin{document}

\begin{flushright}
September 23, 2016 \\
\end{flushright}

\vs{1.5cm}





\begin{center}
{\Large\bf

 Soft See-Saw: Radiative Origin of Neutrino Masses  \\

 \vs{0.2cm}

in SUSY Theories  }
\end{center}

\vspace{0.5cm}
\begin{center}
{\large
{}~Luka Megrelidze$^{\hs{0.5mm}}$\footnote{E-mail: luka.megrelidze.1@iliauni.edu.ge}{}~ and
{}~Zurab Tavartkiladze$^{\hs{0.5mm}}$\footnote{E-mail: zurab.tavartkiladze@gmail.com}
}
\vspace{0.5cm}

{\em Center for Elementary Particle Physics, ITP, Ilia State University, 0162 Tbilisi, Georgia}
\end{center}

\vspace{0.6cm}

\begin{abstract}

Radiative neutrino mass generation within supersymmetric (SUSY) construction is studied.
  The mechanism is considered where the lepton number violation is originating from the soft SUSY breaking terms.
  This requires MSSM extensions with states around the TeV scale.
   We present several explicit realizations based on extensions either by MSSM singlet or  $SU(2)_w$ triplet states.
   Besides some novelties of the proposed scenarios,  various phenomenological implications are also discussed.

\end{abstract}

\hspace{0.4cm}{\it Keywords:}~Lepton number violation; Neutrino masses; supersymmetry; radiative corrections.

\section{Introduction}

One of the missing pieces of the Standard Model (SM) is the consistent neutrino sector required for accommodation
of the neutrino date \cite{nu-data}. Extensions based on type I \cite{Minkowski:1977sc},
 type II \cite{Magg:1980ut} and  type III  \cite{Foot:1988aq}
see-saw mechanisms have been suggested, which at tree level induce effective dimension five ($d=5$) $\De L=2$ lepton number violating
operator \cite{weinb-d5-l2-ops}
\beq
\fr{\lam_{ij}}{M_*}l_il_j HH ~,
\la{d5-LNV2-op}
\eeq
where $l_i$ ($i, j=1,2,3$ are family indices) and $H$ are SM lepton and Higgs doublets respectively. The (\ref{d5-LNV2-op})
 type couplings, in turn, generate neutrino masses and mixings after EW symmetry breaking.

It was shown \cite{Zee:1980ai,Babu:2010vp,Ma:2006km,fras}
that, augmenting the SM by specific states and couplings, the operators of Eq. (\ref{d5-LNV2-op}) can be generated
radiatively at one (or higher) loop level. This possibility, referred to as radiative neutrino mass generation mechanism, offers
many interesting scenarios with rich phenomenological implications \cite{Zee:1980ai,Babu:2010vp,Ma:2006km,fras}.
It is a curious fact that, besides some studies \cite{SUSY-radNUmass}, within SUSY constructions such possibilities have not
been pursued much.\footnote{Note that, like within SM, also in minimal supersymmetric extension of the SM (MSSM),
 it is hard to see how neutrino masses  $\stackrel{>}{_\sim } 0.05$~eV can be generated  even with non-renormalizable [i.e. the cut off scale
 $M_*=M_{Pl}$ in Eq. (\ref{d5-LNV2-op})]  interactions are taken into account.}
  In this work, aiming to fill this gap and address this issue within SUSY scenarios, we offer possibilities of
  loop induced neutrino masses, where lepton number violation  occurs in soft SUSY breaking couplings and is  transferred
  in the SM neutrino sector at the loop level. Referring to this possibility as `soft see-saw' mechanism, we present several extensions
  which naturally realize this program. The following three types of extensions are considered:
  {\it i)} extension with MSSM singlet (matter) superfields,   {\it ii)} extension with a pair of $SU(2)_w$ triplet-antitriplet scalar
  superfields carrying  $U(1)_Y$ hypercharges $\pm 2$, and {\it iii)} extension with matter $SU(2)_w$ triplet and $U(1)_Y$ neutral
  superfields. We call these three scenarios soft type I, soft type II and soft type III see-saw scenarios respectively. We work out
  details of each construction and discuss phenomenological implications.

Note that the radiative neutrino mass generation within MSSM extension by right handed neutrinos and lepton number violation
by soft SUSY breaking terms (the scenario we refer to as soft type I see-saw) has been considered in papers of Ref. \cite{SUSY-radNUmass}.
While in these works variations of models with $\De L=2$ right handed sneutrino couplings have been suggested, we present concise and detailed
 discussion of this setup, derive effective $\De L=2$ operators, outline necessary ingredients and give some constraints.
As far as the soft type II and soft type III see-saw scenarios are concerned, these radiative neutrino mass generation mechanisms are
new. As will be shown, these scenarios have various interesting ingredients and peculiar phenomenological implications.

  The paper is organized as follows. In the next section we discuss lepton number violating  operators of different dimensions,
   involving MSSM states and list some of them. Then, as a demonstration, we compute the $d=5$, $\De L=2$ operator of type (\ref{d5-LNV2-op}),
   induced via 1-loop due to presence of a quartic ($d=4$) $\De L=2$ operator. In Sect. \ref{sect-soft-seesaw}  we present models -  extensions of
   the MSSM - where the lepton number violation takes place in soft SUSY breaking terms and show how integrating out the extra states
   gives effective $\De L=2$ operators. We consider extensions with MSSM singlet matter (right handed neutrino) superfields and
   with $SU(2)_w$ triplet states. Based on these extensions different  scenarios (e.g. type I, type II-A, type II-B and type III soft see-saw)
   emerge. In each case neutrino masses are induced via loops. Corresponding 1-loop and 2-loop results are presented.
   Sect.  \ref{sect-concl} contains discussion and outlook with some prospects for future studies. In appendix
   \ref{susy-br} we present supergravity formalism for calculation of soft SUSY breaking terms emerged from hidden sector
   superpotential and non-minimal K\"ahler potential couplings. We give the details of the SUSY breaking via Polonyi superpotential and specific non-minimal K\"ahler potential - the system insuring for SUSY breaking $X$ superfield $F_X\sim m_{3/2}M_{Pl}$ (where $m_{3/2}$ is
   a gravitino mass) and adequately suppressed value for $\lan X\ran $. Both these are needed for our model building.
   In appendix \ref{eff-ops}  detailed derivations of effective $\De L=2$ quartic couplings, emerging by integration of
   MSSM singlet and $SU(2)_w$ triplet states, are given. In appendix \ref{2-loop-calc} we present the calculation of the 2-loop contribution to the
   neutrino masses.

\section{Some $\De L=2$ Lepton Number Violating Operators with MSSM States}

The lepton sector of the MSSM involves the following $D$ and $F$-term couplings:

\beq
\l l^\dag e^{V}l+e^{c\dag}e^Ve^c\r_D \to \tl l^*\tl Vl+\tl e^{c*}\tl Ve^c+{\rm h.c.}
\la{lept-Dterm}
\eeq
\beq
\l Y_Ele^ch_d\r_F+{\rm h.c.}\to Y_E\l le^ch_d+\tl le^c\tl h_d+l\tl e^c\tl h_d\r +{\rm h.c.}
\la{lept-Fterm}
\eeq
where schematically $V$ indicates all appropriate gauge superfields multiplied by proper generator and gauge coupling.
Without soft SUSY breaking terms, it is possible to define the lepton number ($L$-numbers for $l, e^c$) and also slepton number
($\tl L$ -numbers for $\tl l, \tl e^c$) separately.\footnote{Provided that the gauginos $\tl V_a$ and the higgsinos $\tl h_d, \tl h_u$
carry appropriate lepton and slepton charges.}
Thus, the couplings in (\ref{lept-Dterm}) and (\ref{lept-Fterm}) possess two independent symmetries.
However, including soft SUSY breaking terms - the gaugino masses $\fr{1}{2}M_{\tl V_a}\tl V_a\tl V_a$ -
gives the slepton numbers fixed equal to the corresponding lepton numbers: $\tl L=L$.\footnote{The relation $\tl L=L$  also
emerges if, instead of gaugino masses, the trilinear $A$-terms $A_E\tl l\tl e^ch_d$ are included.}
 Thus,  all components from a given superfield  have  same  lepton number.
Therefore, if for instance, in the soft SUSY breaking sector the lepton number will be broken,
 it will be transferred in the fermion sector via radiative corrections.

In our consideration we assume that the whole Lagrangian respects the matter parity, under which states
transform as $\Phi \to (-1)^{3(B-L)+2S}\Phi$, where $B$ and $L$ are baryon and lepton numbers respectively and $S$ indicates the
spin of the $\Phi $ state. This symmetry insures that no baryon and lepton number violating couplings present at renormalizable level
and LSP is a stable.

Considering higher dimensional operators, in SUSY (with matter parity) the lepton number violation starts from the $F$-term $d=5$ operator
\beq
 \int \!\!d^2\te (lh_u)^2+{\rm h.c.}\to \l (ll) h_uh_u +2(l\tl  h_u) \tl lh_u +(\tl h_u\tl h_u)\tl l\tl l\r +{\rm h.c.}
\la{susy-d5-L2-op}
\eeq
which is the SUSY analog of the  (\ref{d5-LNV2-op}) coupling.
At r.h.s of Eq. (\ref{susy-d5-L2-op}) $l$ denotes SM fermionic lepton and $h_u$ is up type Higgs doublet. Symbols with tildes denote
their superpartners respectively.
The coefficients at r.h.s of Eq. (\ref{susy-d5-L2-op}) are related
due to SUSY. However, in general they can be different especially if some of these couplings emerge via SUSY
breaking terms. Therefore, we write each of them with independent coefficients:
\beq
d=5,~~ \De L=2 ~~{\rm operators}:~~~~\lam_5(l\hs{0.5mm}h_u)^2+
\lam_5'(\tl l\hs{0.5mm}\tl h_u)^2+\lam_5''\hs{0.5mm}\tl l \hs{0.5mm}h_u\hs{0.5mm}l\hs{0.5mm}\tl h_u+{\rm h.c.}
\la{d5-L2-ops}
\eeq
$\lam_5$-type couplings  are directly constrained from the neutrino masses. The
$\lam_5'$ and $\lam_5''$ operators induce $\lam_5$-term via loops in which soft SUSY breaking couplings participate.
Therefore, constraints on $\lam_5'$, $\lam_5''$ will  depend on SUSY spectroscopy.
The $\lam_5'$ operator emerges in one of the model (named as type II-B soft see-saw model) we present in Sect. \ref{subsec-typeII}.

Before going to the higher dimensional operators, note that with  MSSM states one can write  $\De L=2$  quartic ($d=4$) coupling
\beq
d=4, ~~\De L=2~~{\rm operator}:~~~~\lam_4(\tl{l}h_u)^2+{\rm h.c.}
\la{d4}
\eeq
This term is non-supersymmetric operator and to discuss its origin one needs to have some UV completion.
In Sect. \ref{sect-soft-seesaw} we present models, where this operator emerges and compute $\lam_4$ coupling in terms of model parameters.

Let us also give some $d=6$, $\De L=2$ couplings involving states from $l$ and $h_u$ superfields:
\beq
d=6, ~~\De L=2~~{\rm operators}:~~~~\lam_6 (l \hs{0.5mm}\tl h_u)^2+\lam_6' (\tl l\tl l) \hs{0.5mm}g\tl V(\tl h_u h_u)+{\rm h.c.}
\la{d6}
\eeq
where in $\lam_6'$-term under $g\tl V$ we assume $g_1\tl V_1$ or $g_2\tl V_2$ (with appropriate contraction of the $SU(2)_w$ gauge indices).
This coupling  emerges in  type II-B soft see-saw model (discussed  in Sect. \ref{subsec-typeII}).

Other $\De L=2$ couplings can be obtained from the operators in Eqs. (\ref{d5-L2-ops})-(\ref{d6}) by the substitution $h_u\to h_d^\dag $.
Also, the variety of $\De L=2$ operators involving $e^c$ and $\tl e^c$ states can be constructed. The models with their generation would be interesting
to consider, but in this work we pursue only specific constructions.

$\De L=2$ operators, including those we gave above, by radiative corrections will be converted to the operator $\lam_5(l\hs{0.5mm}h_u)^2$
 - responsible for the neutrino mass generation. In the next subsection we give details of the calculation of this $\lam_5$-term
  emerged from the $\lam_4$ coupling of Eq. (\ref{d4}) via gaugino/higgsino loop dressings.

\subsection{$\De L=2$ Quartic Coupling and Neutrino Mass}
\la{d4-op}

\begin{figure}[t]
\begin{center}
\hs{-1cm}
\resizebox{0.45\textwidth}{!}{
 \hs{0.5cm} \vs{0.5cm}\includegraphics{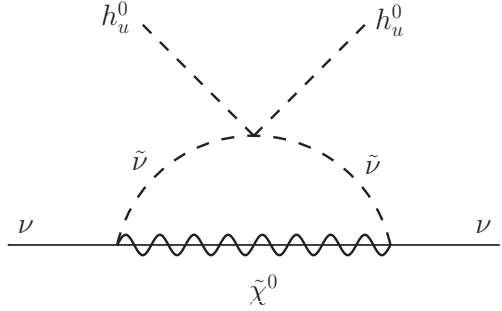}
}
\vs{0.2cm}
\caption{Diagram(s) responsible for neutrino masses via $\lam_4$-coupling.}
\label{fig-d4-to-d5}       
\end{center}
\end{figure}

The coupling (\ref{d4}) is invariant under SM gauge symmetry. Since we are interested in neutrino mass generation, we extract from it
the neutral components:
\beq
\lam_4(\tl{\nu }h_u^0)^2+{\rm h.c.}
\la{lam4}
\eeq
The combination $\tl{\nu }\tl{\nu }$ will be converted to $\nu \nu $ by the neutralino dressing diagram(s) shown in Fig. \ref{fig-d4-to-d5}.
The relevant terms are also
\beq
i\fr{g_1}{2}\tl B\tl l^\dag l-i\fr{g_2}{\sq{2}}\tl{W}_b^a\tl {l_a}^* l^b+{\rm h.c.}\to
\fr{i}{2} \nu \l g_1\tl B -g_2\tl W^0 \r \tl{\nu}^* +{\rm h.c.}
\la{gaugino-l-l}
\eeq
(given in 2-component notations of Ref. \cite{Wess:1992cp}) where $\tl B$ and $\tl{W}_b^a$ are $U(1)_Y$ and $SU(2)_w$ gauginos respectively.
The indices $a, b=1,2$ belong to the $SU(2)_w$ gauge group.
 $\tl W$'s matrix is defined as
\beq
\tl{W}_b^a=
\fr{1}{\sq{2}}\left(\!\!\!
  \begin{array}{cc}
    \tl W^0& \sq{2}\tl W^{+}\\
    \sq{2}\tl W^{-} & -\tl W^0\\
  \end{array}
\!\!\right)^{\!\!a}_{\!\!b}~ .
\la{tlW}
\eeq

We are looking for the  operator:
\beq
\fr{1}{2}\ka^{\nu }_{ij}\nu_i\nu_j (h_u^{0})^2+{\rm h.c.}
\la{d50}
\eeq
where   $\ka^{\nu }_{ij}$ will be computed in terms of couplings appearing in (\ref{lam4}),
(\ref{gaugino-l-l}), together with other model parameters given below.

By diagonalization of $4\tm 4$ neutralino mass matrix (of the states $\{ -i\tl B,  -i\tl W^0 , \tl h_d^0, \tl h_u^0\}$), the  $\tl B$ and  $\tl W^0$  get transformed as
\beq
\tl B =U_{\!\chi 0}^{1\bar a}\hs{0.5mm}\tl \chi^0_{\bar a}~, \hs{1cm}~~~~\tl W^0 =U_{\!\chi 0}^{2\bar a}\hs{0.5mm}\tl \chi^0_{\bar a}~,
~~~~(\bar a=1, 2, 3, 4)~,
\la{Nino-CHino-diag}
\eeq
where $\chi^0_{\bar a}$ is a physical mass eigenstate neutralino with mass $M_{\chi0}^{\bar a}$.
Using these in Eq. (\ref{gaugino-l-l}), the relevant couplings will be
\beq
\fr{i}{2} \nu \l g_1U_{\!\chi 0}^{1\bar a}\hs{0.5mm}-g_2U_{\!\chi 0}^{2\bar a}\hs{0.5mm}\r \chi^0_{\bar a}  \tl{\nu}^*
+{\rm h.c.}
\la{gaugino-nu-l}
\eeq
Interactions in Eqs. (\ref{lam4}), (\ref{gaugino-nu-l}) together with neutralino Majorana type mass terms, generate the operator of Eq. (\ref{d50}) via 1-loop diagram(s) shown in Fig. \ref{fig-d4-to-d5}.
Upon evaluation of the loop integral for the expression of $\ka^{\nu }$ we obtain:
\beq
\ka^{\nu }_{ij}=-\fr{\al_2}{8\pi }(\lam_4)_{ij}
\sum_{\bar a=1}^4\fr{(\tan \te_WU_{\!\chi 0}^{1\bar a}-U_{\!\chi 0}^{2\bar a})^2}{M_{\chi 0}^{\bar a}}
f(x^{i}_{\bar a},x^{j}_{\bar a})~,
\la{kapa5-nu}
\eeq
where $\tan \te_W\!=\fr{g_1}{g_2}$ and
\beq
x^{i}_{\bar a}=\l \!\fr{m_{\tl{\nu }_i }}{M_{\chi0}^{\bar a}}\!\r^{\!\!2} ,~~~
f(x_1,x_2)=\fr{x_1\ln x_1-x_2\ln x_2-x_1x_2\ln \fr{x_1}{x_2}}{(x_1-x_2)(x_1-1)(x_2-1)}~.
\la{V-f}
\eeq
From these expressions we can see that in order to have adequately suppressed neutrino masses($\stackrel{<}{_\sim }0.1$~eV), for
the SUSY particle masses$\sim $~TeV, we need to have $\lam_4\stackrel{<}{_\sim }10^{-8}$. In the next section we will see that this suppression can be
naturally realized within presented models.

\section{Soft See-Saw}
\la{sect-soft-seesaw}
In this section we present models which generate some of the $\De L=2$ operators [in particular: $\lam_4, \lam_5'$ and $\lam_6'$-type
couplings of Eqs. (\ref{d4}), (\ref{d5-L2-ops}) and (\ref{d6}) respectively].
This happens through the soft SUSY breaking sector. Therefore, we refer to it as a soft see-saw mechanism. Scenarios we consider are based on
extensions either with  MSSM singlet matter superfields\footnote{This kind of extension, with lepton number violating  soft SUSY breaking terms,
has been considered in Refs. \cite{SUSY-radNUmass}. Here we give detailed discussion of the setup, necessary ingredients, generation of effective
$\lam_4$ coupling and some constraints.}, or  with a pair of $SU(2)_w$ triplet-antitriplet scalar superfields,
or on a model with matter $SU(2)_w$ triplet superfields. Each case is investigated separately.

\subsection{Type I Soft  See-Saw}
\la{subsec-typeI}
In this case, the MSSM is extended with right handed neutrino (RHN) superfields $N$.
Since within our scenario, the masses of these states are near TeV scale, in order to avoid unacceptably  large neutrino masses,
we will need to suppress the $Y_{\nu } lNh_u$ type Yukawa  superpotential couplings. Instead demanding ad hoc condition $|Y_{\nu }| \stackrel{<}{_\sim }10^{-7}$,
we will forbid such superpotential coupling by the $R$-symmetry, under which the superfield $\phi_i$ and the superpotential
transform as:
\beq
\phi_i\to e^{iR_i} \phi_i~,~~~~~~~~W\to e^{i\om }W~.
\la{R-tr}
\eeq
With the $R$ charge assignment for the superpotential $W$ and superfields
$ l, h_u, N$  given in Table \ref{R-typeI},
 together with the couplings $lNh_u$, also the $NN$-type superpotential terms are forbidden.
 However, as we will see shortly, the lepton number violating soft SUSY breaking terms will be induced.
These type of terms will come from the K\"ahler potential, via the SUSY breaking. The latter
will occur  in a hidden sector with MSSM singlet superfield $X$ having  non-zero $F$-term $F_X\sim m_{3/2}M_{Pl}$. The $R$ charge of the $X$ is selected to be
$R_X=1$.
In Table \ref{R-typeI} we summarize $R$ charges of all superfields to be considered.
%
%
%
%
\begin{table}
\vs{-0.8cm}

 $$\begin{array}{|c|c|c|c|c|c|}

\hline

\vs{-4mm}
&  &  &   && \\
& W  & X  & h_u  & N & l \\
\vs{-5mm}
&  &  &   & &\\

\hline

\vs{-3.5mm}
&  &  &   && \\
R& \om & 1 &  \al & \fr{1}{2} & -\fr{1}{2}-\al \\
\vs{-3.6mm}
&  &  &   && \\

\hline
\hline

\vs{-4mm}
&  &  &   && \\
& h_d  & e^c  & q  & u^c & d^c \\
\vs{-5mm}
&  &  &   & &\\

\hline

\vs{-3mm}
& &  & & &\\
R& 1-\al & \om \!+\!2\al \!-\!\fr{1}{2} &  \al_q & \om \!-\!\al_q\!-\!\al & \om \!-\!\al_q\!+\!\al \!-\!1\\

 \hline
\end{array}$$
\caption{$R$ charges in the type I soft  see-saw model. $\om \neq m+1/2$ ($m\in Z$).
}
 \label{R-typeI}
\end{table}
%
%
With this assignment,
the following minimal and non-minimal K\"ahler  (denoted below as ${\cal K}_{m}$ and ${\cal K}_{nm}$ respectively)  potential couplings are allowed:
\beq
{\cal K}_{m}=\sum_f  f^\dag e^{g_aV_a}f ~,~~~
{\cal K}_{nm}=\fr{X^\dag }{2M_{\rm Pl}}  \ka NN + \fr{X^\dag X}{\bar M^3}\ka_A lNh_u +\fr{X^\dag }{M_{\rm Pl}}\ka_hh_uh_d +{\rm h.c.}
\la{Kal-N}
\eeq
(On the second term in ${\cal K}_{nm}$, with cut off scale $\bar M<M_{\rm Pl}$, we comment below.)
With these, the generated soft SUSY breaking terms involving $\tl N$ will be:
 \beq
V(\tl N)=\tl l^TA_N\tl Nh_u+\fr{1}{2}\tl N^TB_N\tl N +{\rm h.c.}+\tl N^\dag M_{\tl N}^2\tl N~,
\la{soft-BA}
\eeq
where  $\tl l^T=(\tl l_1, \tl l_2, \tl l_3)$ and $\tl{N}^T=(\tl N_1, \tl{N}_2, \cdots )$ (the number of $\tl N$ states should be
$\geq 2$). The matrix $M_{\tl N}^2$  is hermitian while $B_N$ is a symmetric:
$M_{\tl N}^2=(M_{\tl N}^2)^\dag $ and $B_N^{T}=B_N$. The mass$^2$ terms in (\ref{soft-BA}) with
 $F_X\sim m_{3/2}M_{\rm Pl}$ are
$M_{\tl N}^2\sim |F_X|^2/M_{\rm Pl}^2\sim m_{3/2}^2$.
The soft $B$-terms are
 $B_N\sim \ka |F_X|^2/M_{\rm Pl}^2\sim \ka m_{3/2}^2$, while with $\bar M\sim (m_{3/2}M_{Pl}^2)^{1/3}$ the trilinear $A$-term will be
 $A_N\sim \ka_A |F_X|^2/\bar M^3\sim \ka_A m_{3/2}$.
For detailed discussion and derivations  see Appendix \ref{susy-br}. In the same Appendix we discuss the generation of $F_X$ via Polonyi
superpotential and specific  K\" ahler potential for $X$.

From the couplings in Eq. (\ref{soft-BA}), by integrating out the $\tl N$ states we get the operator of Eq. (\ref{d4}).
In particular, by
 assumption $\ka \stackrel{<}{_\sim } 1/3,  \ka_Av_u\stackrel{<}{_\sim }0.3m_{3/2}$, the expression for $\lam_4$ can be approximated as:
 \beq
 \lam_4\simeq \fr{1}{2}A_N\fr{1}{M_{\tl N}^2}B_N^\dag \fr{1}{(M_{\tl N}^2)^T}A_N^T~.
 \la{deriv-lam4}
 \eeq
 Corresponding diagram is given in Fig. \ref{fig-soft-type1}.
 Derivation and more accurate expression for $\lam_4$ is given in Appendix \ref{eff-ops} by Eq. (\ref{exact-lam4-N}).
 From the structure of (\ref{deriv-lam4}) and process given in Fig. \ref{fig-soft-type1} it is clear why it is fair to call this
 mechanism (for generating of $\lam_4$ coupling)
 the type I soft  see-saw. The quartic operator of (\ref{d4}) is forbidden in the SUSY limit, but is generated due to SUSY breaking $\De L=2$ terms
 via integrating out the $\tl N$ states. With these, as was shown in  section \ref{d4-op}, the neutrino masses are generated at 1-loop level.
The needed suppression for the $\lam_4$ is easily obtained.  For instance, with the SUSY particle masses $\sim $ few TeV
and  $\fr{A_N}{M_{\tl N}}\sim \fr{\sq{B_N}}{M_{\tl N}}\sim 10^{-2}$, from (\ref{deriv-lam4})
we can get $\lam_4\sim 10^{-8}$ - guaranteeing suppressed values of the neutrino masses($\sim 0.1$~eV).
\begin{figure}[t]
\begin{center}
\hs{-1cm}
\resizebox{0.6\textwidth}{!}{
 \hs{0.5cm} \vs{0.5cm}\includegraphics{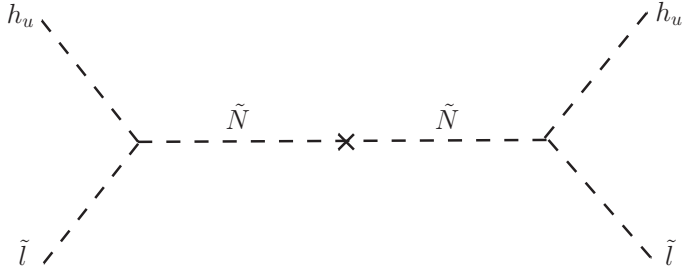}
}
\vs{0.2cm}
\caption{Diagram generating $(\tl lh_u)^2$ term via $\tl N$ exchange.}
\label{fig-soft-type1}       
\end{center}
\end{figure}

The second operator in ${\cal K}_{nm}$ of Eq. (\ref{Kal-N}) can be obtained by integrating some heavy states. Here we give one example.
By introducing additional MSSM singlet states ${\cal N}, \ov{\cal N}, {\cal N}'$ and $\ov{\cal N}'$ we can have the K\"ahler
coupling ${\cal K}_{{\cal N}}=\fr{\ka_{{\cal N}}X^\dag}{M_{Pl}}N\ov{\cal N}$ and the superpotential terms:
$W_{{\cal N}}=M_{{\cal N}}{\cal N}\ov{\cal N}+\lam_XX{\cal N}\ov{\cal N}'+{M'}_{{\cal N}}{\cal N}'\ov{\cal N}'+\lam_{{\cal N}}\ov{\cal N}'lh_u$.
One can easily verify that integration of the states ${\cal N}, \ov{\cal N}, {\cal N}',\ov{\cal N}'$
induces the  K\"ahler  operator $\fr{\ka_{{\cal N}}\lam_X\lam_{{\cal N}}}{M_{Pl}M_{{\cal N}}{M'}_{{\cal N}}}X^\dag XlNh_u$.
Comparing this with Eq. (\ref{Kal-N}), one can identify
$\fr{k_A}{\bar M^3}=\fr{\ka_{{\cal N}}\lam_X\lam_{{\cal N}}}{M_{Pl}M_{{\cal N}}{M'}_{{\cal N}}}$.

As shown in Appendix  \ref{susy-br}, by the construction one can insure that the VEV of the $X$ field can be
adequately suppressed.  With $\lan X\ran \stackrel{<}{_\sim }10^{-7}M_{Pl}$ the neutrino Dirac Yukawa couplings, generated via the
   K\"ahler potential will be $Y_{\nu}\stackrel{<}{_\sim }10^{-7}$ (see expression in (\ref{Yuk-pol}) and related discussion before and after of this equation),
   i.e. not relevant for the neutrino masses.
On the other hand, there is a low bound on the values of $Y_{\nu}$ couplings. They should be sizable enough to
insure decays of the fermionic RHN states $N$ within $\stackrel{<}{_\sim }0.3$~sec. in order to not affect the standard Big Bang nucleosynthesis.
With $M_N>M_l+M_{h_u} (M_{\tl l}+M_{\tl h})$ we will have decays $ N\to lh_u(\tl l \tl h)$, with the lifetime given by
\beq
\Ga_{N\to lh_u(\tl l \tl h)}^{-1}\approx \fr{8\pi }{|Y_{\nu }|^2M_N}=0.3~{\rm sec.}\tm
 \l \fr{2.34\cdot 10^{-13}}{|Y_{\nu}|}\r^2 \l \fr{1 ~{\rm TeV}}{M_N}\r .
\la{N-lifetime}
\eeq
Since $Y_{\nu }\sim \lan X\ran/M_{Pl}$, we will have the low bound on the VEV $\lan X\ran \stackrel{>}{_\sim }5.6\cdot 10^{5}$~GeV.
Therefore, summarizing all above, we will have the following range:
\beq
2.34\cdot 10^{-13}\stackrel{<}{_\sim }Y_{\nu }\stackrel{<}{_\sim } 10^{-7}~~~~\lrarr ~~~~
5.6\cdot 10^{5}~{\rm GeV}\stackrel{<}{_\sim } \lan X\ran \stackrel{<}{_\sim } 2\cdot 10^{11}~{\rm GeV}.
\la{Yuk-X-range}
 \eeq
The mass $M_N$ is generated from the first coupling of ${\cal K}_{nm}$
 of (\ref{Kal-N}), with the value
 $M_N\sim \ka F_X/M_{\rm Pl}\sim \ka m_{3/2}$.  Note that, it is easy to satisfy  $M_N>M_l+M_{h_u} (M_{\tl l}+M_{\tl h})$
 insuring the decays described in (\ref{N-lifetime}).

Closing this subsection,  with the $R$-charge assignment given in Table \ref{R-typeI},
 the MSSM Yukawa superpotential couplings are:
\beq
W_Y^{MSSM}=Y_Uqu^ch_u+Y_Dqd^ch_d+Y_Ele^ch_d~.
\la{Yuk-MSSM}
\eeq
 Note that the direct $h_uh_d$ superpotential term is forbidden.
However, as shown first in Ref. \cite{Giudice:1988yz}, the last coupling in  (\ref{Kal-N})  generates $\mu \sim \ka_hm_{3/2}$ and $B\mu \sim \ka_hm_{3/2}^2$
terms.

With the $R$-charge assignments given in Table \ref{R-typeI} the lepton number violating superpotential couplings $X^nh_ul$ (which break also matter
parity)  are all forbidden
for $\om \neq m+1/2$ (where $m$ is an integer). Also the trilinear lepton number violating interactions $qd^cl$ and $e^cll$ will be forbidden in the superpotential.
 The baryon number violating term $u^cd^cd^c$ can be forbidden
if the phases $\al_q$, $\om $ and $\al $ will  satisfy additional condition $3\al_q-2\om-\al \neq m_1$ ($m_1 \in Z $).
With these and with one more condition $3\al_q-\om-\al \neq m_2$ ($m_2 \in Z $)
the $d=5$ baryon and lepton number violating couplings  $qqql$, $u^cu^cd^ce^c$ and $qqqh_d$ will be automatically forbidden.
 One can also make sure that with these three conditions
\beq
\om \neq m+1/2~,~~~3\al_q-2\om-\al \neq m_1~, ~~~3\al_q-\om-\al \neq m_2~~~ ~~(m, m_{1,2} \in Z )~,
\la{conds-phases}
\eeq
  baryon and matter parity violating couplings do not emerge also from the K\"ahler  potential.

\subsection{Type II Soft  See-Saw}
\la{subsec-typeII}

In this case we extend the MSSM by introducing the pair of $SU(2)_w$ triplet and antitriplet superfields $\Si $, $\bar{\Si }$ with $U(1)_Y$ hypercharges
$2$ and $-2$ respectively (in this normalization, the hypercharge of the lepton doublet $l=(e^-, \nu )$ equals to $-1$). $\Si$ and $\bar{\Si }$ are $2\tm 2$
matrices in $SU(2)_w$ space and their compositions are given by:
\beq
\Si_{ab}\!=\!\left(
           \begin{array}{cc}
             \Si^{(++)} & \Si^{(+)}/\sq{2} \\
              \Si^{(+)}/\sq{2}  &  \Si^{(0)}  \\
           \end{array}
         \right)_{\!\!ab}~,~~~~
         \bar{\Si }^{ab}\!=\!\left(
           \begin{array}{cc}
             \bar{\Si}^{(--)} & \bar{\Si}^{(-)}/\sq{2} \\
              \bar{\Si}^{(-)}/\sq{2}  &  \bar{\Si}^{(0)}  \\
           \end{array}
         \right)^{\!\!ab}~,~
\la{tr-antitr-matr}
\eeq
where signs in superscripts indicate electric charges [e.g. $(++)$ stands with the double charged state].

With this field content, two scenarios which somewhat differ from each other, can be considered. We refer to them as
type II-A soft  and type II-B soft see-saw models.

\vs{0.2cm}
{\bf Type II-A soft  see-saw model}
\vs{0.1cm}

\begin{figure}[t]
\begin{center}
\hs{0.9cm}
\resizebox{0.95\textwidth}{!}{
 \hs{0.5cm} \vs{0.5cm}\includegraphics{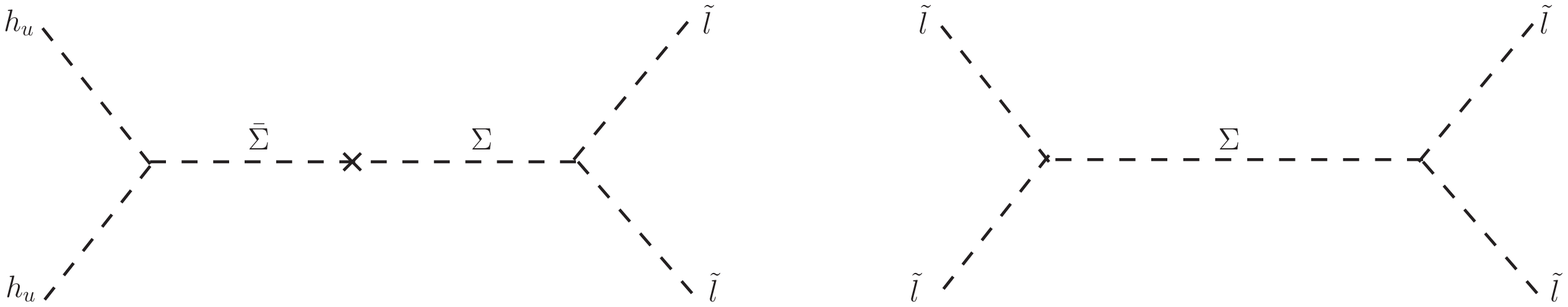}
}
\vs{0.2cm}
\caption{Diagrams generating $(\tl lh_u)^2$ (left) and $(\tl l\hs{0.3mm}\tl l)(\tl l^{\hs{0.3mm}*}\tl l^{\hs{0.3mm}*})$ (right) terms by
integration of scalar triplets $\Si, \bar{\Si }$.}
\label{fig-soft-type2}       
\end{center}
\end{figure}

\hs{-0.6cm}In this case,  the $R$-charge assignment is given in Table \ref{R-typeII-A}
%
%
%
%
%
\begin{table}
\vs{-0.8cm}

 $$\begin{array}{|c|c|c|c|c|c|}

\hline

\vs{-4mm}
&  &  &   && \\
& X & h_u  & \Si  & \bar{\Si } & l \\
\vs{-5mm}
&  &  &   & &\\

\hline

\vs{-3.5mm}
&  &  &   && \\
R& 1 & \al &  1+2\al & -2\al & -\fr{1}{2}-\al \\
\vs{-3.6mm}
&  &  &   && \\

\hline
\hline

\vs{-4mm}
&  &  &   && \\
& h_d  & e^c  & q  & u^c & d^c \\
\vs{-5mm}
&  &  &   & &\\

\hline

\vs{-3mm}
& &  & & &\\
R& 1-\al & \om \!+\!2\al \!-\!\fr{1}{2} &  \al_q & \om \!-\!\al_q\!-\!\al & \om \!-\!\al_q\!+\!\al \!-\!1\\

 \hline
\end{array}$$
\caption{$R$ charges in the type II-A soft  see-saw model, with $R_W=\om $. By $\om  \neq m+1/2$ ($m\in Z$)
the matter parity is automatic.
}
 \label{R-typeII-A}
\end{table}
%
%
%
and  relevant K\" ahler potential couplings are:
\beq
{\cal K}_{m}=\sum_f  f^\dag e^{g_aV_a}f ~,~~~
{\cal K}_{nm}=\fr{\ka }{M_{\rm Pl}} X^\dag \bar{\Si }\Si + \fr{X^\dag X}{\bar M^3}  \!\l \!\ka_{A\bar{\Si}} \bar{\Si}h_uh_u
+\fr{1}{2}\ka_{A\Si}\Si ll\r \!+\fr{\ka_h }{M_{\rm Pl}} X^\dag h_uh_d +{\rm h.c.}
\la{Kal-Si}
\eeq
The MSSM superpotential couplings will be same as given in (\ref{Yuk-MSSM}).
The couplings of Eq. (\ref{Kal-Si}) and SUSY breaking hidden sector, together with the $\mu$ and $B$-terms for the
MSSM Higgs doublets, generate the following  soft terms involving the scalar components of $\Si$ and $\bar{\Si}$:
\beq
V(\Si,\bar{\Si})=B_{\Si}\bar{\Si}\Si +A_{\bar{\Si}}\bar{\Si}h_uh_u+\Si \hs{0.4mm}\tl l^{\hs{0.4mm}T}A_{\Si}\tl l+{\rm h.c}
+M_{\Si}^2\Si^\dag \Si +M_{\bar{\Si}}^2\bar{\Si}^\dag \bar{\Si}~.
\la{soft-Si}
\eeq
 Here $\Si $ and $\bar{\Si}$  denote bosonic components. Their fermionic partners will be denoted by $\tl{\Si}$ and $\tl{\bar \Si}$ respectively.
Integration of $\Si, \bar{\Si}$ states leads to the operator of Eq. (\ref{d4}) with approximate expression for $\lam_4$ given by:
\beq
\lam_4\simeq \fr{1}{M_{\Si}^2M_{\bar{\Si}}^2}A_{\bar{\Si}}B_{\Si}^*A_{\Si}~,
\la{lam4-Si}
\eeq
valid for $M_{\Si}M_{\bar{\Si}}\stackrel{>}{_\sim }|B_{\Si}|/3$.
More accurate expression is given in Appendix \ref{eff-ops} [see Eq.  (\ref{exact-type2-d4})].
The first diagram of Fig. \ref{fig-soft-type2} corresponds to the generation of this operator, which gives loop induced neutrino masses as shown
in Sect. \ref{d4-op}. Also within this case, the needed suppression of $\lam_4$ can be achieved, for example, by
the selection $\fr{A_{\Si}}{M_{\Si}}\sim \fr{A_{\bar{\Si}}}{M_{\Si}}\sim \fr{\sq{B_{\Si}}}{M_{\bar{\Si}}}\sim 10^{-2}$.

Via exchange of $\Si, \bar{\Si }$  states, the four slepton interaction
operator is also generated, which in the same approximation (as Eq. (\ref{lam4-Si})) is:
\beq
{\cal L}_{4\tl l}^{\rm eff}=\fr{1}{M_{\Si}^2}\l \tl l^{\hs{0.4mm}T}_aA_{\Si }\tl l_b\r \l \tl l^{\hs{0.4mm}\dag}_aA_{\Si }^*\tl l^{\hs{0.3mm}*}_b \r,
\la{4slept-ASi}
\eeq
where $a,b=1,2$ are $SU(2)_w$ indices [see Eq. (\ref{exact-type2-4l}) for more accurate expression].
The corresponding diagram is the second one in Fig. \ref{fig-soft-type2}.
\begin{figure}[t]
\begin{center}
\hs{0.9cm}
\resizebox{0.45\textwidth}{!}{
 \hs{0.5cm} \vs{0.5cm}\includegraphics{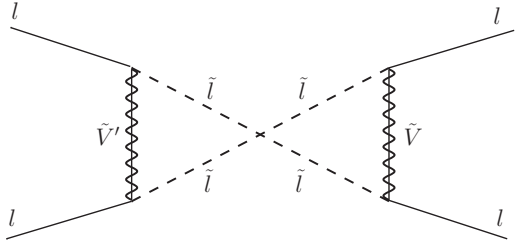}
}
\vs{0.2cm}
\caption{Diagram generating 4-fermion operator via the effective coupling of Eq. (\ref{4slept-ASi}).}
\label{fig-4-ferm-op}       
\end{center}
\end{figure}
By 2-loop gaugino/higgsino dressing diagrams, the operator (\ref{4slept-ASi}) will be converted to the four lepton operator
$C^{4l}_{ijmn}(l^i_al^j_b)(\bar l^m_a\bar l^n_b)$. One relevant diagram is shown in Fig. \ref{fig-4-ferm-op}.
These couplings (i.e. $C^{4l}_{ijmn}$) in turn induce
$e_i\to e_je_me_n$ rare decays, including processes such as $\tau \to 3\mu$, $\mu \to 3e$ etc.
With all SUSY particles and $\Si , \bar{\Si}$ states
having the common masses$\approx M_S$ we estimate $C^{4l}_{ijmn}\approx \l \fr{\al_2}{4\pi}\r^2\fr{(A_{\Si })_{ij}(A_{\Si }^*)_{mn}}{M_S^4}$.
 For instance, for the branching ratio of the reaction  $\mu \to 3e$  we will have
\beq
{ Br}(\mu \to 3e)\approx \l \fr{\al_2}{4\pi}\r^4\fr{|(A_{\Si })_{11}(A_{\Si }^*)_{12}|^2}{G_F^2M_S^8}\simeq
3.8\cdot 10^{-13}\l\fr{1~{\rm TeV}}{M_S}\r^4\left | \fr{(A_{\Si })_{11}(A_{\Si }^*)_{12}}{M_S^2}\right |^2~.
\la{Br-mu-3e}
\eeq
We see that with $M_S\sim 1$~TeV, the current experimental limit  ${ Br}(\mu \to 3e)<10^{-12}$ \cite{Beringer:1900zz}
is easily satisfied even with $\fr{(A_{\Si})_{11}}{M_S}, \fr{(A_{\Si})_{12}}{M_S}\sim 1$. If the latter ratios are taken to be suppressed,
then one can allow to have $M_S$'s values below the TeV scale. Since the experimental limits on $\tau $'s rare decays are less stringent \cite{Beringer:1900zz},
it is easier to satisfy
bounds on branching ratios $Br(\tau \to e_ie_je_k)$. Since $A_{\Si }$  couplings also enter in the neutrino mass matrix [see Eq. (\ref{lam4-Si})],
it would be interesting to investigate their flavor structure in connection to the neutrino data and the rare lepton decays.
  These will open window to probe the neutrino mass generation mechanism presented here.

Finally, the $\ka $-term in Eq. (\ref{Kal-Si}) generates the mass term for the fermionic $\tl{\Si}, \tl{\bar \Si}$  components $\mu_{\Si} \tl{\bar \Si}\tl{\Si}$ with $\mu_{\Si}\sim \ka m_{3/2}$.

\vs{0.2cm}
{\bf Type II-B soft  see-saw model}
\vs{0.1cm}

\hs{-0.6cm}In this modified version, the neutrino masses are induced at 2-loop level. The $R$-charges of the states are given in Table \ref{R-typeII-B}
%
%
%
%
%
\begin{table}
\vs{-0.3cm}

 $$\begin{array}{|c|c|c|c|c|c|}

\hline

\vs{-4mm}
&  &  &   && \\
& X & h_u  & \Si  & \bar{\Si } & l \\
\vs{-5mm}
&  &  &   & &\\

\hline

\vs{-3.5mm}
&  &  &   && \\
R& 1 & \al &  1+2\al -\om & \om -2\al & \fr{\om }{2}-\fr{1}{2}-\al \\
\vs{-3.6mm}
&  &  &   && \\

\hline
\hline

\vs{-4mm}
&  &  &   && \\
& h_d  & e^c  & q  & u^c & d^c \\
\vs{-5mm}
&  &  &   & &\\

\hline

\vs{-3mm}
& &  & & &\\
R& 1-\al & \fr{\om }{2}\!+\!2\al \!-\!\fr{1}{2} &  \al_q & \om \!-\!\al_q\!-\!\al & \om \!-\!\al_q\!+\!\al \!-\!1\\

 \hline
\end{array}$$
\caption{$R$ charges in the type II-B soft  see-saw model, with $R_W=\om $. By $\om \neq 2m+1$ ($m\in Z$)
the matter parity is automatic.
}
 \label{R-typeII-B}
\end{table}
%
%
%
and the K\" ahler potential couplings are
\beq
{\cal K}_{m}=\sum_f  f^\dag e^{g_aV_a}f ~,~~~
{\cal K}_{nm}=\fr{\ka }{M_{\rm Pl}} X^\dag \bar{\Si }\Si +
\fr{X^\dag X}{2\bar M^3}  \ka_{A\Si}\Si \hs{0.4mm}ll \!+\fr{\ka_h }{M_{\rm Pl}} X^\dag h_uh_d +{\rm h.c.}
\la{Kal-BSi}
\eeq
 The $X^\dag \bar{\Si }h_uh_u$ term is forbidden in the K\" ahler potential, but the following superpotential coupling, involving $\bar{\Si}$, is allowed:
\beq
W_{\bar{\Si }}=\lam_{\bar{\Si }}\bar{\Si}h_uh_u~.
\la{sup-BSi}
\eeq
The MSSM superpotential terms are same as given in  Eq. (\ref{Yuk-MSSM}).
\begin{figure}[t]
\begin{center}
\hs{-1cm}
\resizebox{0.8\textwidth}{!}{
 \hs{0.5cm} \vs{0.5cm}\includegraphics{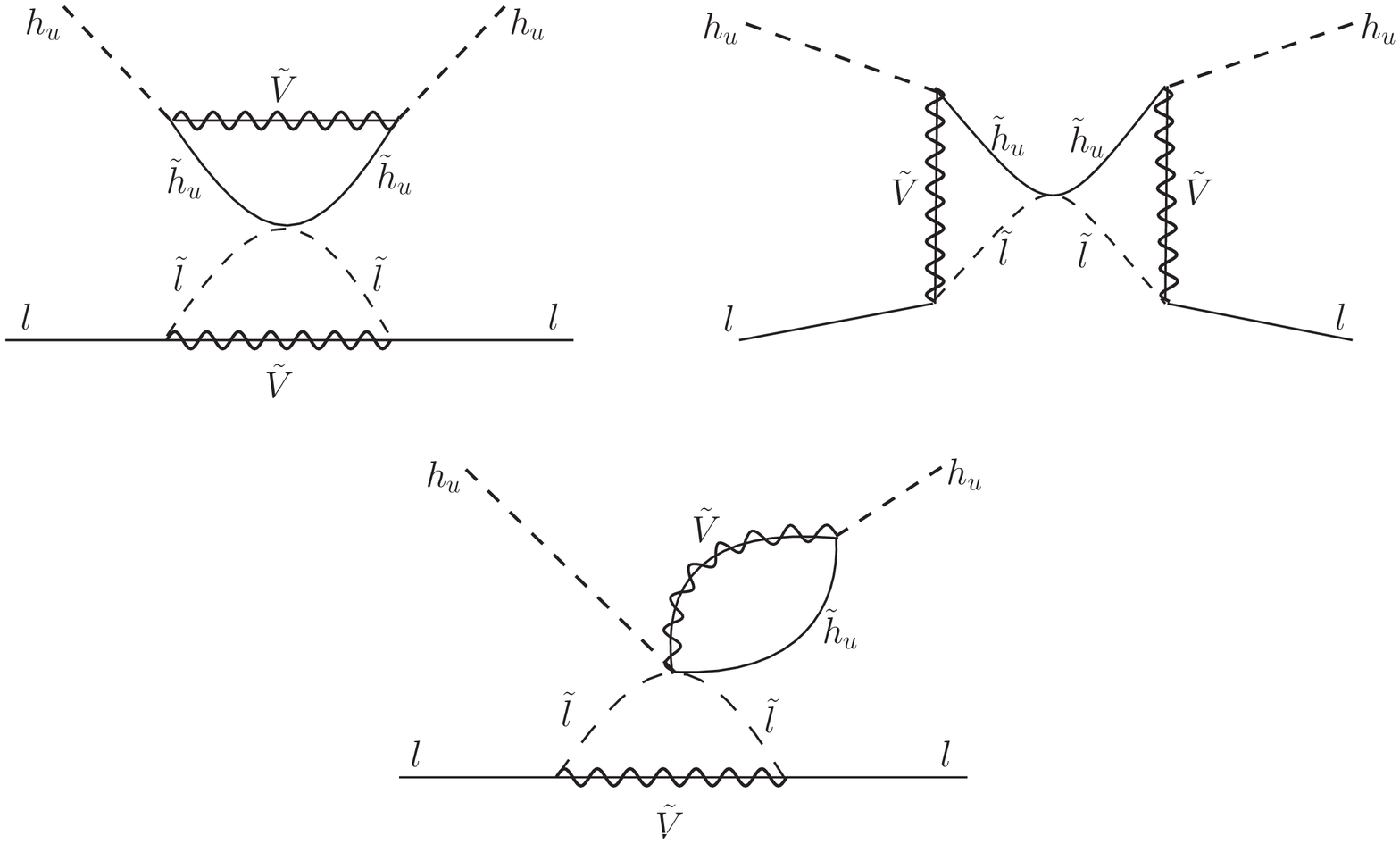}
}
\vs{0.2cm}
\caption{2-Loop diagrams responsible for neutrino masses within  type II-B soft see-saw model.}
\label{fig-2loop1}       
\end{center}
\end{figure}
The coupling (\ref{sup-BSi}), in general would generate the trilinear soft term
$A_{\bar{\Si }}\bar{\Si}h_uh_u$.  However, within the SUSY breaking
scenario we are considering, this $A$-term is suppressed:
\beq
A_{\bar{\Si }}\sim \lam_{\bar{\Si }}m_{3/2}\fr{\lan X\ran }{M_{Pl}}\stackrel{<}{_\sim }10^{-7}\lam_{\bar{\Si }}m_{3/2}~.
\la{ASifromW}
\eeq
(see Eq. (\ref{A-lam-term}) and discussion therein).

K\" ahler potential terms of Eq. (\ref{Kal-BSi}), besides higgsino $\mu$ and Higgs $B$ terms induce the potential terms:
\beq
B_{\Si}\bar{\Si}\Si +A_{\Si}\Si \tl l\tl l+{\rm h.c}+
M_{\Si}^2\Si^\dag \Si +M_{\bar{\Si}}^2\bar{\Si}^\dag \bar{\Si}~.
\la{soft-BSi}
\eeq
With the couplings in (\ref{ASifromW}), (\ref{soft-BSi}), integration of $\Si, \bar{\Si}$ states leads to the operator of Eq. (\ref{d4}) with
expression for $\lam_4$ having the same form as given in Eq. (\ref{lam4-Si}), but with extra strong suppression factor$\stackrel{<}{_\sim }10^{-7}$.
Therefore, 1-loop contribution to the neutrino masses will be very suppressed and can be ignored. Neutrino masses will be induced at 2-loop level,
which we discuss below.

The superpotential of Eq. (\ref{sup-BSi}) gives the following Yukawa interactions:
\beq
\int d^2\te W_{\bar{\Si }}\to \lam_{\bar{\Si }}\l \bar{\Si}\tl h_u\tl h_u+2\tl{\bar{\Si}}\tl h_u h_u\r~.
\la{BSi-sup-to-Yuk}
\eeq
These couplings, together with those given in (\ref{soft-BSi}), upon integration of scalar  components
from ${\bar{\Si}}, \Si $ superfields, induce the following $\De L=2$ dimension $5$ operator ($\lam_5'$-type of Eq. (\ref{d5-L2-ops})):
\beq
\lam_5'(\tl l\hs{0.5mm}\tl h_u)^2~,~~~~{\rm with}~~~~\lam_5'\simeq \fr{1}{M_{\Si}^2M_{\bar{\Si}}^2}\lam_{\bar{\Si}}B_{\Si}^*A_{\Si}~.
\la{d5pr-BSi}
\eeq
The way of derivation of this $d=5$ operator is similar to that given in Appendix \ref{eff-ops}. We just need
to make replacement $A_{\bar{\Si}}h_uh_u\to \lam_{\bar{\Si}}\tl h_u\tl h_u$ in Eq. (\ref{exact-type2-d4}).
The operator (\ref{d5pr-BSi}), by the gaugino dressings at 2-loop gives the $d=5$ operator $\lam_5(l\hs{0.5mm} h_u)^2$ responsible for the neutrino mass.
The relevant diagrams are given in Fig. \ref{fig-2loop1}. For consistency, one should also take into account the effective $d=6$, $\De L=2$ operators
\beq
-\fr{2}{M_{\Si}^2\mu_{\Si}}\lam_{\bar{\Si}}(\tl l^{\hs{0.4mm}T} A_{\Si}\tl l)(g\tl V\tl h_uh_u)+{\rm h.c.}
\la{eff-coup-BSi}
\eeq
(the $\lam_6'$-type coupling of Eq. (\ref{d6})) which are induced by integration of the scalar and fermionic components
from ${\bar{\Si}}, \Si $ superfields [by using
 the mass term $\mu_{\Si }\tl{\bar{\Si}}\tl{\Si}$ and couplings (\ref{soft-BSi}), (\ref{BSi-sup-to-Yuk})].
In (\ref{eff-coup-BSi}), $g\tl V=\{g_1\tl B, g_2\tl W \}$ denote $U(1)_Y$ and/or $SU(2)_w$ gaugino contributions.

\begin{figure}[t]
\begin{center}
\hs{-1cm}
\resizebox{1.01\textwidth}{!}{
 \hs{5cm} \vs{0.5cm}\includegraphics{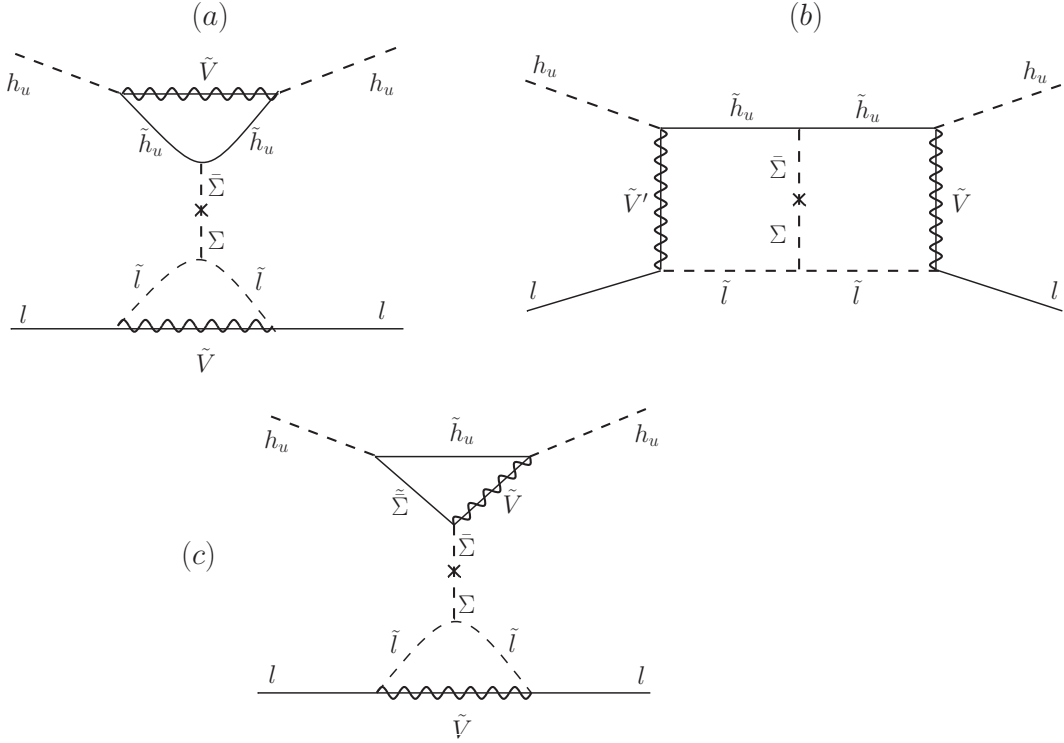}
}
\vs{0.2cm}
\caption{2-Loop diagrams responsible for neutrino masses within  type II-B soft see-saw model (opening effective vortexes).}
\label{fig-2loop2}       
\end{center}
\end{figure}

Care must be taken to treat properly divergences of the diagrams in Fig.  \ref{fig-2loop1}. If the effective vortexes (\ref{d5pr-BSi}) and
(\ref{eff-coup-BSi}) are used without specifying the model (the operators are induced from), the loops need to be cut off by the characteristic scale.
Since within  type II soft see-saw, these operators are obtained by integrating $\Si , \bar{\Si}$ states, full calculation can be done. With 'opening'
 vortexes, the  diagrams are shown in Fig. \ref{fig-2loop2}.
 It is obvious that
log divergence of the first and third diagrams are coming from  upper triangles corresponding to the parts of one loop renormalization
of the soft $A_{\bar{\Si}}$ term (vanishing at tree level). Indeed,
$$
\de A_{\bar{\Si}}=\de A_{\bar{\Si}}^{(1)}+\de A_{\bar{\Si}}^{(2)}~,~~~~~{\rm with}
$$
$$
\de A_{\bar{\Si}}^{(1)}=-\fr{\lam_{\bar{\Si}}}{16\pi^2}\!\sum_{a=1}^3 \!c^{(1)}_ag^2_aM_{\tl{V}_a}
\!\!\l \!\ln \fr{\La^2}{M_{\tl{V}_a}^2}+\fr{\mu^2(2M_{\tl{V}_a}^2-\mu^2)}{(M_{\tl{V}_a}^2-\mu^2)^2}\ln \fr{\mu^2}{M_{\tl{V}_a}^2}
-\fr{\mu^2}{\mu^2-M_{\tl{V}_a}^2} \!\r ,
$$
$$
\de A_{\bar{\Si}}^{(2)}=-\fr{\lam_{\bar{\Si}}}{16\pi^2}\!\sum_{a=1}^3 \!c^{(2)}_ag^2_aM_{\tl{V}_a}
\!\!\l \!\ln \fr{\La^2}{M_{\tl{V}_a}^2}-\fr{\mu^4 \ln \fr{\mu^2}{M_{\tl{V}_a}^2}}{(\mu^2-\mu_{\Si}^2)(\mu^2-M_{\tl{V}_a}^2)}
- \fr{\mu_{\Si }^4 \ln \fr{\mu_{\Si }^2}{M_{\tl{V}_a}^2}}{(\mu_{\Si }^2-\mu^2)(\mu_{\Si}^2-M_{\tl{V}_a}^2)}\!\r ,
$$
\beq
{\rm with}~~~~~~c^{(1)}_a=\l \fr{6}{5}, 3, 0\r ,~~~~c^{(2)}_a=\l \fr{3}{5}, 6, 0\r ,
\la{A-renorm}
\eeq
where $\mu $ and $\mu_{\Si }$ are the MSSM higgsino $\mu$-term and the mass of the fermionic  $\tl{\Si}, \tl{\bar{\Si}}$ states respectively,
  $g_a$ is the gauge coupling (with $c_a^{(1,2)}$ being corresponding group theoretical factors and $M_{\tl{V}_a}$  the corresponding gaugino's soft mass).
  In the last line of (\ref{A-renorm}), $a=1,2,3$ labels $U(1)_Y$, $SU(2)_w$ and $SU(3)_c$ gauge groups respectively (the coupling $g_1$ is taken in $SU(5)$ normalization).
The $\ln \La $ terms coincide with those obtained  via integration of the RG equation for the $A_{\bar{\Si}} $ soft term.
Expressions (\ref{A-renorm}), after the renormalization, leave us with a non-divergent and finite parts.
Assuming the boundary condition $A_{\bar{\Si}}(M_{UV})=0$ at some scale $M_{UV}$ (which can be the Planck mass, in case of gravity mediated SUSY breaking),
at the SUSY scale $M_{S}$, we will have:
\beq
{\rm at}~~~~\mu=M_{S}:~~~~~~~A_{\bar{\Si}}^{\rm eff}=-\fr{\lam_{\bar{\Si}}}{16\pi^2}\!\l \!\sum_{a=1}^3 \!c_ag^2_aM_{\tl{V}_a}\!\r
\!\ln \fr{M_{UV}^2}{M_S^2} , ~~~~c_a=\l \fr{9}{5}, 9, 0\r .
\la{A-eff}
\eeq
Further, we compute the lower triangle diagrams of Fig. \ref{fig-2loop2}-(a),(c), which by the gaugino dressings convert $\tl l\tl l$ to $ll$.
Then, integrate out the $\Si, \bar{\Si}$ states and obtain the contribution to the neutrino mass to be given by:\footnote{Here we ignore the EW symmetry breaking effects, e.g. the gaugino-higgsino mixings.}
\beq
\l \de M_{\nu}^{ij}\r_{\rm Fig. 5-(a)+(c)}=\fr{v_u^2A_{\bar{\Si}}^{\rm eff}}{16\pi^2}\fr{B_{\Si}^*A_{\Si}^{ij}}{M^2_{\Si}M^2_{\bar{\Si}}}
\sum_{a=1}^3 \!\fr{c^{(1)}_ag^2_a}{M_{\tl{V}_a}}f\!\left [\!\l \!\fr{m_i}{M_{\tl{V}_a}}\!\r^2\!\!\!, \l \!\fr{m_j}{M_{\tl{V}_a}}\!\r^2\right ]~,
\la{2-loop-Mnu-5ac}
\eeq
where the function $f$ is defined in Eq. (\ref{V-f}) and $m_i, m_j$ denote masses of $\tl l_i, \tl l_j$.

There is no any divergence from the  diagram of Fig. \ref{fig-2loop2}-(b).
Detailed evaluation of this diagram is given in Appendix \ref{2-loop-calc}. Contribution to the neutrino mass matrix from the 2-loop diagram of
  Fig. \ref{fig-2loop2}-(b) is given by:
\beq
\l \de M_{\nu}^{ij}\r_{\rm Fig. 5-(b)}=\fr{v_u^2}{2}
\mu^{2} g^{2}{g'}^2 A_{\Sigma}^{ij} B_{\Sigma}^{*} \lambda_{\bar{\Sigma}}~J~,
\la{2-loop-Mnu}
\eeq
where $J$ is given in Eq. (\ref{final-J}) and $\gamma _{\alpha nm}$ and other appearing factors are defined in
Eqs. (\ref{simpl-J-1}), (\ref{fracs1}) and (\ref{AnBm})
respectively.

 Let us give an estimate of constraints on the parameters required to obtain the correct suppression of neutrino masses within this scenario.
For simplicity, assume that all SUSY particle masses and $M_{\Si}, M_{\bar{\Si}}, \mu $ are same and equal to $M_S\sim 1$~TeV.
 With this, the expressions at the r.h.s of Eqs. (\ref{2-loop-Mnu-5ac}) and (\ref{2-loop-Mnu}) are simplified to be
$\approx 27 \lam_{\bar{\Si}}\l \fr{\al_2}{4\pi }\r^2\fr{v_u^2B_{\Si}^*A_{\Si}^{ij}}{M_S^4}\ln \fr{M_{Pl}^2}{M_S^2}$ and
$\approx 4\cdot 10^{-3} \lam_{\bar{\Si}}\l \fr{\al_2}{4\pi }\r^2\fr{v_u^2B_{\Si}^*A_{\Si}^{ij}}{M_S^4}$ respectively.
Taking into account these, for $M_S\simeq 1$~TeV and with the selection $\lam_{\bar{\Si}}\sim \fr{A_{\Si}}{M_{\Si}}\sim \fr{\sq{B_{\Si}}}{M_{\bar{\Si}}}\sim 4\cdot 10^{-3}$ we obtain the needed value of $\l \de M_{\nu}^{ij}\r_{\rm Fig. 5-(a)+(c)}\sim 0.1$~eV, while
$\l \de M_{\nu}^{ij}\r_{\rm Fig. 5-(b)} \sim 2\cdot 10^{-7}$~eV is strongly suppressed. As we see, for this choice of the parameters, the
contribution from (\ref{2-loop-Mnu-5ac}) dominates over the one given in Eq. (\ref{2-loop-Mnu}). However, with different spectroscopy these
contributions may be comparable and require detailed investigation, which should be performed elsewhere.

Concluding, note that additional operators where $h_u$ scalar is replaced by $h_d^\dag$ will be also induced. The relevance of this kind of operators
will depend on the value of the VEV $\lan h_d^{(0)}\ran$. The latter, on the other hand, depends on the parameter $\tan \bt$.
 Thus, the  details and numerical results will depend on SUSY spectroscopy, which should be taken such that LHC constraints are satisfied.
  The latter study is beyond the scope of this work.

\subsection{Type III Soft See-Saw}
\la{subsec-typeIII}

In this case, the MSSM is extended with $SU(2)_w$ triplets superfields $T$ with zero hypercharges.
For a realistic neutrino sector, at least two such superfields should be introduced.
 The matrix representation of this superfield(s) has
the form
\beq
T^a_b=\left(
        \begin{array}{cc}
          T^{(0)}/\sq{2} & T^{(-)} \\
          T^{(+)} &-T^{(0)}/\sq{2} \\
        \end{array}
      \right)^{\!\!a}_{\!b}~,
\la{T-matr}
\eeq
with superscripts indicating the electric charges of the fragments.
In this case, the $R$ charges are:
$$
R_W=\om ,~~~R_X=1,~~~R_{h_u}=\al ,~~~R_{T}=\fr{1}{2} ,~~~R_l=-\fr{1}{2}-\al ,~~~R_{h_d}=1-\al ,
$$
\beq
R_{e^c}=\om +2\al-\fr{1}{2},~~~R_q=\al_q,~~~R_{u^c}=\om -\al-\al_q,~~~R_{d^c}=\om -\al_q+\al-1 .
\la{type3-R-ch}
\eeq
Note that with $\om  \neq m+\fr{1}{2}~(m\in Z)$ the matter parity is automatic.
The superpotential couplings with the $T$ states will be forbidden, while the
 K\" ahler potential couplings will be
\beq
{\cal K}_{m}=\sum_f  f^\dag e^{g_aV_a}f ~,~~~
{\cal K}_{nm}=\fr{\ka }{2M_{\rm Pl}} X^\dag TT  + \fr{X^\dag X }{\bar M^3} \ka_A lTh_u
\!+\fr{\ka_h }{M_{\rm Pl}} X^\dag h_uh_d +{\rm h.c.}
\la{Kal-T}
\eeq
The $R$ charge selection of Eq. (\ref{type3-R-ch}) is consistent with MSSM superpotential couplings given in Eq. (\ref{Yuk-MSSM}).
From the coupling of (\ref{Kal-T}), likewise of previous cases,  together with the MSSM  $\mu$ and $B$-terms,  the
following soft SUSY breaking terms are induced:
\beq
\fr{1}{2}T^TB_{T}T +\tl l^TA_{T}Th_u+{\rm h.c}+T^\dag M_{T}^2T~,
\la{soft-T}
\eeq
where by $T$ the scalar components are denoted. Their fermionic partners $\tl T$ gain the mass $\fr{1}{2}\mu_T\tl T\tl T$ with $\mu_T\sim \ka m_{3/2}$.
(Once more, we refer the reader to Appendix \ref{susy-br} for details.)
Now, one can easily verify that integration of $T$'s scalar components, via (\ref{soft-T}) couplings   induce  the $\De L=2$
operator of Eq. (\ref{d4}) with:
\beq
\lam_4\simeq \fr{1}{2}A_T\fr{1}{M_{T}^2}B_T^\dag \fr{1}{M_{T}^2}A_T^T~.
\la{lam4-T}
\eeq
This expression is approximate and works well with $M_T^2\stackrel{>}{_\sim }B_T/3$.
The derivation of this expression is very similar to that corresponding to the case of type I soft see-saw model, presented in Appendix \ref{eff-ops}.\footnote{
For obtaining (\ref{lam4-T}) one can do  the replacements $B_N\to B_T$ and $A_N\to A_T$ in Eq. (\ref{exact-lam4-N}) and then use the proper approximation.}
With the $\lam_4$ coupling the neutrino mass will be generated (as discussed in Sect. \ref{d4-op}) at 1-loop level via diagram shown in Fig. \ref{fig-d4-to-d5}.
1-Loop contribution will dominate if
  $Y_Tl\tl Th_u$ type Yukawa interactions, induced through the K\"ahler potential (with $Y_T\sim \lan X\ran /M_{Pl}$) will be adequately suppressed.
This requires $\lan X\ran \stackrel{<}{_\sim }2\cdot 10^{11}$~GeV. For this case [unlike the Eq. (\ref{Yuk-X-range})] we will not have low bound on $\lan X\ran $,
because the  states from $T$ can decay via MSSM gauge interactions.

From the potential (\ref{soft-T}), upon integration of the  $T$ states,  between $\tl l$ and $T$, the mixing of the order $\sim A_T\fr{v_u}{M_T^2}$ is emerged.
This, in turn, via $T$-gaugino (higgsino) loops will induce $e_i\to e_j\ga $  decays. However, with $M_T\sim $ few TeV  and $A_T/M_T\stackrel{<}{_\sim }1/10$,
 the $\tl l-T$  mixing will be$\stackrel{<}{_\sim }10^{-2}$. This will be enough for adequate suppression of such rare processes.
Much will also depend on details of the flavor structure of $A_T$ and $M_T^2$ matrices, which require separate investigation and is not pursued here.

\vs{0.3cm}

Concluding, let us emphasize that
the decays of the states from $\Si, \ov{\Si}$ and $T$ (within type II and type III  soft see-saw scenarios respectively) in SM and LSP states
proceed via EW interactions and there is no low bound on the value of $\lan X\ran $ (unlike to the type I soft see-saw scenario;
 see the discussion before and after Eq. (\ref{N-lifetime})).

\section{Discussion and Outlook}
\la{sect-concl}

In this paper we have presented possibilities for radiative neutrino mass generation within SUSY theories. Suggested
mechanisms open broad prospects for further investigations.

Since extensions, we have suggested, have the lepton number violation in the soft SUSY breaking sector with new states near the
TeV scale, the models would have peculiar collider  signatures.  This gives possibilities for testing the origin of neutrino masses
at accelerator experiments. This issue deserves separate investigation in a spirit of Refs. \cite{seesaw-LHC}, \cite{Babu:2009aq}.
It would be also interesting to exercise the lepton number violating higher dimensional operators (such as $d\geq 7$, $\De L=2$ couplings
studied in different scenarios earlier \cite{highDim-L2-ops}, \cite{Babu:2009aq})
and investigate how they may emerge from the soft SUSY breaking terms.

Since within presented models new states lie near the TeV scale and they have some family dependent interactions, one expects
to have new contributions to the rare processes such as $\mu \to e\ga, \tau \to (e, \mu )\ga, \mu \to 3e$, etc.
These also could be the signatures of the presented scenarios and may serve for models' test.
Similar concern to new possible contributions to the EW precision parameters ($T, S$ and $U$).

 Finally, would be challenging to embed considered models in SUSY Grand Unification (GUT) such as $SU(5)$ and $SO(10)$ GUTs.
 Because of the GUT symmetry, new relations and constraints would emerge, making models more predictive.
 These and related issues will be addressed elsewhere.

\subsubsection*{Acknowledgments}
We thank K.S. Babu, B. Bajc and M. Nemevsek for discussions and helpful comments.
The work is partially supported by Shota Rustaveli National Science Foundation (Contracts No. 31/89 and No. DI/12/6-200/13).
 Z.T. thanks CETUP* (Center for Theoretical Underground Physics and Related Areas)
 for its hospitality and partial support during 2015 and 2016 Summer Programs.
 Z.T. also would like to express a special thanks to the Mainz Institute for Theoretical Physics (MITP) for its hospitality and support.

\appendix

\renewcommand{\theequation}{A.\arabic{equation}}\setcounter{equation}{0}

\section{SUSY Breaking.  $\mu $, $B$ and $A$-type Couplings}
\la{susy-br}

In the superconformal formulation, the supergravity Lagrangian density is given by \cite{conf-sugra}, \cite{Kugo:1982mr}
$$
{\cal L}_{D}+{\cal L}_F~~~~~~~~{\rm with}
$$
\beq
{\cal L}_{D}=- 3\int \!\!\!d^4\theta \,e^{-{\cal K}/(3M_{Pl}^2)}\phi^+\phi ,   ~~~~~~~~~~~~~
{\cal L}_F=\fr{1}{M_{Pl}^3}\int \!\!\!d^2\theta\phi^3\, W +\frac{1}{4}\int \!\!\!d^2\te f_{IJ}{{\cal W}}^{\alpha I}{{\cal W}}_{\alpha}^J+ {\rm h.c.}
\label{eq:eleven}
\eeq
where  ${\cal K}$ and $W(\Phi)$  are   the K\" ahler potential and the superpotential respectively,
while $f_{IJ}(\Phi)$ is the gauge kinetic function (for the chiral superfield strength ${{\cal W}}^{\alpha I}$ obtained from the vector
 superfield $V^I$). $\phi$ is the compensator chiral superfield.
 Superspace integrals of (\ref{eq:eleven}) (and throughout this paper)
should be understood as F and D-densities of (conformal) SUGRA as given in Refs. \cite{conf-sugra}, \cite{Kugo:1982mr}.

In general, the scalar potential
\beq
V=V_F+V_D~,
\la{tot-V}
\eeq
consists of two parts. The $D$-term potential
\beq
V_D=\fr{1}{2}\sum_ag_a^2D_a^2 ~,
\la{Dterm-pot}
\eeq
and the $F$-term potential $V_F$. The latter is obtained from (\ref{eq:eleven}) by
integrating  $F$-terms. Doing so, setting $\phi =M_{Pl}$  and going to the  Einstein-frame one obtain:
\be
 V_F=\,e^{{\cal K}/M_{Pl}^2}\left(D_{\bar J}{\bar W}{\cal K}^{{\bar J}I}D_IW - \fr{3}{M_{Pl}^2}|W|^2\right),
 ~~~ {\rm with }~~~ D_I\equiv\partial_I+\fr{1}{M_{Pl}^2}{\cal K}_I~.
\label{eq:formula}
\ee
A few comments about the definitions are in order.
In the 'covariant' derivative $D_I$ (with respect to the field $\Phi_I$) we have:
$\partial_I\equiv \fr{\pl }{\pl \Phi_I}$ and ${\cal K}_I\equiv \fr{\pl {\cal K}}{\pl \Phi_I}$. The object
 ${\cal K}^{{\bar J}I}$ is an inverse of the matrix ${\cal K}_{I{\bar J}}$ build from the second derivatives of the   K\" ahler
potential  ${\cal K}_{I{\bar J}}=\fr{\pl^2 {\cal K}}{\pl \Phi_I \pl \Phi_I^\dag }$. Thus,
$$
{\cal K}_{I{\bar M}}{\cal K}^{{\bar M}J}=\de_I^J~,~~~~~~{\cal K}^{{\bar I}M}{\cal K}_{M{\bar J}}=\de^{\bar I}_{\bar J}~.
$$
The $DW$ can be defined as a column, while $D\bar W$ is its hermitian conjugate row:
$$
D\bar W=\l (D_1W)^*, (D_2W)^*,\cdots \r ~.
$$
Thus, the first entry in the brackets of Eq. (\ref{eq:formula}) can be written as $D\bar W({\cal K}'')^{-1}DW$.

We will be dealing with superfields  $X_{i}$ of the visible sector and with  superfield $X$ through which the SUSY  breaking takes place
 in a hidden sector. For all chiral superfields  unified notation $\Phi_I=\{ X, X_i\}$ is used.
The K\" ahler potential ${\cal K}$, together with minimal quadratic (kinetic) terms, will also include non-holomorphic higher order terms:
\beq
{\cal K}=\sum_I \Phi_I^\dag e^{g_aV_a}\Phi_I+\Phi_I^\dag \l \!\ka_{(2)}^{IJK}\Phi_J\Phi_K+\ka_{(3)}^{IJKM}\Phi_J\Phi_K\Phi_M+\cdots \r
+\cdots +{\rm h.c.}
\la{kal-terms}
\eeq
The superpotential
\beq
W=W_v(X_i)+W_h(X)
\la{tot-sup}
\eeq
is the sum of the visible and hidden sector superpotentials, denoted by $W_v(X_i)$ and  $W_h(X)$  respectively.
The form of $W_h(X)$ should insure SUSY breaking, i.e. $F_X\neq 0$. Within our consideration,
the couplings in Eq. (\ref{kal-terms}), in combination with the superpotential, will be responsible for generation of $\mu $ type terms as well as
 for $B$ and $A$-type soft SUSY breaking terms.
Namely, with the expression of the scalar potential  we can calculate the mass$^2$  terms (corresponding to soft mass$^2$ and $B$-terms) and trilinear
scalar couplings (corresponding to the soft SUSY breaking $A$-terms):
\beq
m^2_{i\bar j}=\fr{\pl^2 V}{\pl X_i\pl X_j^\dag }~,~~~~~B_{ij}=\fr{\pl^2 V}{\pl X_i\pl X_j}~,~~~~~A_{ijk}=\fr{\pl^3 V}{\pl X_i\pl X_j\pl X_k}~.
\la{soft-terms}
\eeq
More detailed discussion about these couplings will be given in the next subsection after discussing the details of the SUSY breaking.

\subsection{SUSY Breaking via Polonyi Superpotential}

For the SUSY breaking hidden sector superpotential $W_h(X)$ we consider Polonyi superpotential  \cite{Polonyi:1977pj} of the form:
\beq
W_h=JX +C~.
\la{Wh-1}
\eeq
The coupling $J$ insures non-zero $F_X$, while the constant $C$ is needed to cancel cosmological constant.
The hidden sector superpotential (as usually) explicitly breaks the $R$-symmetry.\footnote{This is desirable to avoid cosmological
difficulties with massless $R$-axion, emerging from the spontaneous breaking of the continuous $R$-symmetry.
  For interesting interconnection  between $R$-symmetry and SUSY breaking  see Ref. \cite{Nelson:1993nf}.}
 This may be considered as a soft breaking, because
it does not makes troubles in a visible sector.
With $X$'s minimal K\" ahler potential  ${\cal K}=X^\dag X $, the superpotential (\ref{Wh-1}), by proper selection of $J$ and $C$ gives SUSY breaking
Minkowski vacuum, but with $\lan X\ran \sim M_{Pl}$.
 The reason for this is the following. Minimum condition in a Minkowski vacuum  fixes $J$ and $C$. With this, the potential for $X$ has
 the linear term $\sim m_{3/2}^2M_{Pl}X$ and the quadratic term $\sim m_{3/2}^2|X|^2$ (i.e. potential's curvature)
and the VEV of $X$ is basically set from the ratio of these two terms: $\lan X\ran \sim m_{3/2}^2M_{Pl}/m_{3/2}^2=M_{Pl}$.

The situation will change, i.e. one can get desirable value of $F_X$ and suppressed
$\lan X\ran $, with  higher order terms in the  K\" ahler potential. For our purposes will be enough to include the coupling
$\De {\cal K}(X)=-(X^\dag X)^2/(4{M_*}^2)$. As we will see, with $M_* \ll M_{Pl}$ the scalar potential will develop high curvature($\sim m_{3/2}^2\fr{M_{Pl}^2}{M_*^2}$), which will suppress the VEV of $X$.
Thus, we will consider $X$'s K\" ahler potential:
\beq
{\cal K}(X)=X^\dag X -\fr{1}{4{M_*}^2}(X^\dag X)^2
\la{X-kal-1}
\eeq
and analyze this case in details. Since the gravitino mass is given by \cite{Wess:1992cp}, \cite{Freedman:2012zz}:
\beq
m_{3/2}=e^{\lan {\cal K}\ran/(2M_{Pl}^2)}\fr{\lan W\ran }{M_{Pl}^2} ,
\la{gravitino-mass}
\eeq
and we are looking for a solution with suppressed $\lan X\ran $, we will use the parametrization
 \beq
J=\bar mM_{Pl}~,~~~~C=c\bar mM_{Pl}^2~,~~~~~\al=M_{Pl}/M_* ,~~~~x=X/M_{Pl}~,
\la{param}
\eeq
where $\bar m$ will turn out to be of the order of $m_{3/2}$ and $c$ is a constant. With these, using in (\ref{eq:formula}) forms
of (\ref{Wh-1}) and (\ref{X-kal-1}), we will get the potential:
\beq
V=\bar m^2M_{Pl}^2e^{\cal K}\l \fr{1}{1-\al^2|x|^2}\left |1+x^* (1-\fr{\al^2}{2}|x|^2)(x+c)\right |^2-3\left |x+c\right |^2 \r
\la{pol-pot-X4kal}
\eeq
One can easily check that in the region $|x|<1/\al $, with $\al \gg 1$ (i.e.  for $M_*\ll M_{Pl}$ ), the potential has the minimum. The constant $c$ can be selected in such a way
that the cosmological constant is zero in this minimum. Doing this, upon the minimization of (\ref{pol-pot-X4kal}), we  find:
\beq
\lan X\ran =M_{Pl}\lan x\ran \simeq \fr{2}{\sq{3}}\fr{M_*^2}{M_{Pl}}\l 1-\fr{11}{3}\l \fr{M_*}{M_{Pl}}\r^2+\fr{74}{3}\l \fr{M_*}{M_{Pl}}\r^4+\cdots \r~,~
\la{x-vev}
\eeq
\beq
c\simeq \fr{1}{\sq{3}}\l 1-\fr{2}{3}\l \fr{M_*}{M_{Pl}}\r^2+\fr{14}{9}\l \fr{M_*}{M_{Pl}}\r^4 -\fr{190}{27}\l \fr{M_*}{M_{Pl}}\r^6+\cdots \r .
\la{cosm-c}
\eeq
In finding these we have used an expansion with powers of $M_*/M_{Pl}$. So, the solution can be found up to the needed accuracy keeping appropriate power of  $M_*/M_{Pl}$.
With (\ref{x-vev}), (\ref{cosm-c}) and (\ref{param}), from Eq. (\ref{gravitino-mass}) we find:
\beq
\bar m\simeq \sq{3}m_{3/2}.
\la{m32-barm}
\eeq
In the minimum, the masses of real and imaginary  scalar components of the $X$ field will be $\simeq \sq{3}m_{3/2}M_{Pl}/M_*$.
The region $|x|<1/\al $  has minimum with a SUSY breaking.
Although, the VEV of the $X$ field can be strongly suppressed, the $F_X$-term and  $D_XW$ in the minimum are of the order of $\sim m_{3/2}M_{Pl}$:
\beq
\lan F_X\ran \simeq \sq{3}m_{3/2}M_{Pl}~,~~~~~~~~~~~~\lan D_XW\ran =\lan \pl_XW+\fr{{\cal K}_X}{M_{Pl}^2}W\ran \simeq \sq{3}m_{3/2}M_{Pl} .
\la{FX-DXW-X4kal}
\eeq
 In the regions $|x|>/\al $  the potential is not bounded from below. However, the tunneling from the branch of $|x|<1/\al $  to the either other branches are extremely suppressed.
As far as the K\" ahler coupling $-(X^\dag X)^2/(4{M_*}^2)$ [the second term of Eq. (\ref{X-kal-1})] is concerned, it can be generated
by integration of states of mass
$\sim M_*$ which couple with the $X$ field. For instance, having two chiral superfields $\Om$, $\ov{\Om}$ and the superpotential couplings
$\lam X\Om^2+M_{\Om}\Om \ov{\Om}$, at 1-loop level the   K\" ahler potential receives the quartic correction\footnote{Simple way for computing the loop corrections to the K\" ahler potential
is to use formalism given in Ref. \cite{Brignole:2000kg}.}
 $\De {\cal K}(X)=-\fr{|\lam |^4}{12\pi^2M_{\Om}^2}(X^\dag X)^2$.
This justifies analysis performed above.

Within considered framework,  the visible sector superfields $X_i$ will couple with $X$ via higher order terms in the K\" ahler potential.
 Within our consideration, the latter  will have the form:
\beq
{\cal K}(X_i)=X_i^\dag X_i+\fr{X^\dag X}{6\bar M^3}\ka_A^{ijk}X_iX_jX_k+\fr{X^\dag }{2M_{Pl}}\ka^{ij}X_iX_j+{\rm h.c.}
\la{Kel-Xi-pol}
\eeq
These couplings, as shown in \cite{Giudice:1988yz}  will generate $\mu $,  $B$ and $A$-type terms, as well as
 the Yukawa couplings. From the  expression of the $F$-term  potential (\ref{eq:formula}), we have the following relevant terms:
\beq
-\fr{|F_X|^2}{6\bar M^3}\ka_A^{ijk}X_iX_jX_k-\fr{F_XW^*}{2M_{Pl}^3}\ka^{ij}X_iX_j+{\rm h.c.}=-\fr{m_{3/2}^2M_{Pl}^2}{2\bar M^3}\ka_A^{ijk}X_iX_jX_k-\fr{\sq{3}}{2}m_{3/2}^2\ka^{ij}X_iX_j+{\rm h.c.}
\la{A-B-pol}
\eeq
Therefore we have obtained the $B$-terms:
 \beq
 B^{ij}=-\sq{3}m_{3/2}^2\ka^{ij} ,
 \la{B-pol}
 \eeq
  of the needed value. The $A$-terms will have desirable values with $\bar M=(3m_{3/2}M_{Pl}^2)^{1/3}\sim 10^{-5}M_{Pl}$.
With this scale, we will have
\beq
A^{ijk}=-m_{3/2}\ka_A^{ijk} ,~~~~{\rm with}~~~~\bar M=(3m_{3/2}M_{Pl}^2)^{1/3} .
\la{A-pol}
\eeq

For the fermionic states $\Psi_i$ (coming from $X_i$ superfield) the $\mu $-terms $\fr{1}{2}\mu^{ij}\Psi_i\Psi_j$
 can be calculated from the mass formulae \cite{Freedman:2012zz}:
\beq
M_{ij}=\fr{m_{3/2}M_{Pl}^2}{\lan W\ran }\l (\pl_{X_i}+\fr{{\cal K}_{X_i}}{M_{Pl}^2})\nabla_{X_j}W-\Ga^I_{X_iX_j}\nabla_IW\r -\fr{3}{2}\fr{M_{Pl}^2}{\lan W\ran^2}(\nabla_{X_i}W)(\nabla_{X_j}W)~.
\la{Yuk-kal-pol}
\eeq
Using this and the forms of (\ref{Wh-1}), (\ref{X-kal-1}) and (\ref{Kel-Xi-pol}), we get:
\beq
\mu^{ij}=-\sqrt{3}m_{3/2}\ka^{ij}~.
\la{mu-pol-pot}
\eeq
From the mass formulae  (\ref{Yuk-kal-pol}), we can also extract the Yukawa couplings. With the definition of the Yukawa coupling $\fr{1}{2}Y^{ijk}X_i\Psi_j\Psi_k$,
 with $\nabla_{X_i}W=\pl_{X_i}W$ (valid within this model) and using (\ref{Kel-Xi-pol}), (\ref{A-pol}), we obtain
\beq
Y^{ijk}=\fr{2A^{ijk}}{\sq{3}m_{3/2}}\fr{\lan X\ran }{M_{Pl}}=-\fr{2}{\sq{3}}\ka_A^{ijk}\fr{\lan X\ran }{M_{Pl}}\simeq -\fr{4}{3}\l \fr{M_*}{M_{Pl}}\r^2\ka_A^{ijk}~.
\la{Yuk-pol}
\eeq
Since, $\lan X\ran $ can be strongly suppressed, the Yukawa coupling also can have desirable suppression. Namely, with $\fr{M_*}{M_{Pl}}\stackrel{<}{_\sim}3\cdot 10^{-4}$,
if the $Y^{ijk}$ corresponds to the neutrino Dirac Yukawa couplings $Y_{\nu}$, then according to Eq. (\ref{Yuk-pol})
they will be  $|Y_{\nu}|\stackrel{<}{_\sim}10^{-7}$, i.e. irrelevant for the neutrino masses.

If the (visible) superpotential involves the trilinear Yukawa-type terms
\beq
W^{(3)}=\fr{1}{6}\lam^{ijk}X_iX_jX_k~,
\la{W-tril}
\eeq
then, using (\ref{eq:formula}) and (\ref{soft-terms})  we can see that the corresponding $A$-term is also induced due to $\lam^{ijk}$ coupling,
but is strongly suppressed by the $M_*^2/M_{Pl}^2$ factor:
\beq
A_{\lam}^{ijk}\simeq \fr{1}{3}\l\fr{M_*}{M_{Pl}}\r^2m_{3/2}\lam^{ijk} ~.
\la{A-lam-term}
\eeq

We close this appendix by commenting on the possibility of the soft gaugino mass generation. Since within our framework we are
applying the $R$-symmetry, the corresponding operator should be consistent with it. For instance, if the superpotential's $R$ charge
is selected as $\om =4$,\footnote{With this selection no phenomenologically harmful coupling will be allowed
within models we are considering [see Tables \ref{R-typeI}, \ref{R-typeII-A}, \ref{R-typeII-B} and Eq. (\ref{type3-R-ch})].} then we also have $R({{\cal W}}^{\alpha }{{\cal W}}_{\alpha})=\om =4$ and the $D$-term effective operator responsible for gaugino mass will be
$\int d^4\te \l\fr{X^\dag }{X}\r^2\fr{{{\cal W}}^{\alpha }{{\cal W}}_{\alpha}}{M_{Pl}}$. From this, the gaugino mass will be
$M_{\tl V}\sim \fr{1}{M_{Pl}}\l \fr{X^\dag }{X}\r^2_D=-\fr{2|F_X|^2}{M_{Pl}\lan X\ran^2}\simeq -m_{3/2}\fr{27m_{3/2}}{8M_{Pl}(M_*/M_{Pl})^4}$, which for
$M_*/M_{Pl}\sim 2\cdot 10^{-4}$, $m_{3/2}\sim $~TeV, gives the desirable value $M_{\tl V}\sim m_{3/2}$.
Different possibility, for gaugino mass($\sim m_{3/2}$) generation
would be to include the operator $\int d^2\te \fr{X}{M_{Pl}} {{\cal W}}^{\alpha }{{\cal W}}_{\alpha}$. The latter, although
explicitly breaks $R$-symmetry (similar to hidden sector superpotential), but do not spoil any phenomenology, also can be considered
as a plausible option.

\renewcommand{\theequation}{B.\arabic{equation}}\setcounter{equation}{0}

\section{Deriving Effective Couplings}
\la{eff-ops}

\begin{center}
{\bf $\lam_4$ Coupling induced by $\tl N$ states}
\end{center}

Here we derive the effective operator (\ref{d4}) obtained by integrating out the states $\tl{N}$. The relevant
terms are given in the potential of Eq. (\ref{soft-BA}). Not taking into account the EW symmetry breaking effects, we set $D$-terms to
zero and do not consider their effects.

Since we are deriving an effective operator, relevant at low energies, we ignore the kinetic terms (setting momenta to zero).
Then the equations of motion for $\tl N, \tl N^*$ are:
\begin{equation}
\left.
\begin{array}{c}
-\frac{\partial {\cal L}(\tl N)}{\partial \tilde{N}_{p}^{\ast }}=\fr{\pl V(\tl N)}{\pl \tl{N}_p^*}=
(M_{\tl N}^2)^{pj}\tilde{N}_{j}+A_N^{*\alpha p}\tilde{l}_{\alpha }^{\ast }h_{u}^{\ast }+
B_N^{*pj}\tilde{N}_{j}^{\ast } =0 \\
-\frac{\partial {\cal L}(\tl N)}{\partial \tilde{N}_{p}}=\fr{\pl V(\tl N)}{\pl \tl{N}_p}=
(M_{\tl N}^2)^{jp}\tilde{N}_{j}^*
+A_N^{\alpha p}\tilde{l}_{\alpha }h_{u}+B_{pj}\tilde{N}_{j}=0 ,%
\end{array}%
\right.
\la{eqs-Ntl}
\end{equation}
where $p, j$ indices numerate RHN states, while $\al(=1,2,3)$ is a family index.
We will be interested  in the case when the mass scales $M_{\tl N}^2$ are larger than the terms $A_N\lan h_u^{(0)}\ran$.
Then the mixings between $\tl N$ and $\tl{\nu }$ can be ignored at the leading order.
Therefore,  $\tl N$ states can be integrated from Eqs. (\ref{eqs-Ntl}).  The latter in a matrix form are:
\begin{equation}
\left(
\begin{array}{cc}
M_{\tl N}^2 & B_N^\dag \\
B_N & (M_{\tl N}^2)^{T}%
\end{array}%
\right) \left(
\begin{array}{c}
\tilde{N} \\
\tilde{N}^{\ast }%
\end{array}%
\right) =\left(
\begin{array}{c}
-A_N^\dag \tilde{l}^*h_{u}^{\ast } \\
-A_N^{T}\tilde{l}h_{u}%
\end{array}%
\right)~,
\la{N-eom-inMatrix}
\end{equation}
where the matrix at l.h.s of (\ref{N-eom-inMatrix}) contains the block sub-matrices.
Having in general the block matrix of the form:
\begin{equation}
{\cal M}=\left(
\begin{array}{cc}
A & B \\
C & D%
\end{array}%
\right)~,
\end{equation}
its inverse (in case ${\rm Det} A\neq 0$ and ${\rm Det} D\neq 0$ ) is given by \cite{Wess:1992cp}:
\begin{equation}
{\cal M}^{-1}=\left(
\begin{array}{cc}
\left( A-BD^{-1}C\right) ^{-1} & -A^{-1}B\left( D-CA^{-1}B\right) ^{-1} \\
-D^{-1}C\left( A-BD^{-1}C\right) ^{-1} & \left( D-CA^{-1}B\right) ^{-1}%
\end{array}%
\right)~.
\la{inverse-M}
\end{equation}
Using this, from (\ref{N-eom-inMatrix}) we obtain
\begin{equation}
\tilde{N}\!\!=\!-\!\l \!M_{\tl N}^2-\!B_N^\dag \left( (M_{\tl N}^2)^{T}\r^{\!\!-1}\!B_N\!\r^{-1}\!\!A_N^\dag \tilde{l}^*h_{u}^{\ast }
+(M_{\tl N}^2)^{-1}B_N^\dag \!\left( (M_{\tl N}^2)^{T}-B_N(M_{\tl N}^2)^{-1}B_N^\dag \right) ^{-1}\!A_N^{T}\tilde{l}h_{u}
\la{sol-tilN}
\end{equation}
Plugging the solution (\ref{sol-tilN}) back in (\ref{soft-BA}), after grouping various terms and some simplifications, we obtain the
effective $d=4$, $\De L=2$ interaction term
\beq
{\cal L}_{\rm eff}^{\De L=2}=\tl l^T\lam_4\tl l h_uh_u +{\rm h.c.}
 \la{eff-d4-L2-N}
 \eeq
 i.e. that given in Eq. (\ref{d4}), with:
\beq
\lam_4=\fr{1}{2}A_N(M_{\tl N}^2)^{-1}B_N^\dag \!\left( (M_{\tl N}^2)^{T}-B_N(M_{\tl N}^2)^{-1}B_N^\dag \right) ^{-1}\!A_N^{T}~.
\la{exact-lam4-N}
\eeq
The corresponding diagram is given in Fig. \ref{fig-soft-type1}.
In the case of relatively small $B_N$-terms (e.g. $M_{\tl N}^2\stackrel{>}{_\sim}B_N/3)$, the (\ref{exact-lam4-N}) is simplified to
the form given in Eq. (\ref{deriv-lam4}).

\begin{center}
{\bf $\lam_4$ Coupling and ${\cal L}^{4\tl l}_{\rm eff}$ induced by $\Si$ and $\bar{\Si}$ states}
\end{center}

In this case,   we integrate out the scalar components from the superfields $\Si $ and $\bar{\Si}$. The latter will be denoted by same symbols as
superfields they are coming from. The relevant potential terms are given in Eq. (\ref{soft-Si}). As for the case with $\tl N$ states, here we
also ignore kinetic terms, EW symmetry breaking effects and also mixings of components of $h_u, h_d$ with corresponding states of $\Si ,\bar{\Si}$.
Equations of motion for $\Si^* $ and $\bar{\Si}$ (only two of them are independent) are:
$$
-\fr{\pl {\cal L}}{\pl \Si^*}=\fr{\pl V(\Si, \bar{\Si})}{\pl \Si^*}
=M_{\Si}^2\Si +B_{\Si}^*\bar{\Si}^*+\tl l^{\hs{0.4mm}\dag}A_{\Si}^*\tl l^{\hs{0.4mm}*}=0
$$
\beq
-\fr{\pl {\cal L}}{\pl \bar{\Si}}=\fr{\pl V(\Si, \bar{\Si})}{\pl \bar{\Si}}=
M_{\bar{\Si}}^2\bar{\Si}^{\hs{0.4mm}*}+B_{\Si}\Si+A_{\bar{\Si}}h_uh_u=0~.
\la{Si-barSi-eqs}
\eeq
From (\ref{Si-barSi-eqs}) we find solutions:
\beq
\Si \!=\!\frac{1}{M^{2}_{\Si}M_{\bar{\Si}}^2\!-\!|B_{\Si}|^2}\!\l \!A_{\bar{\Si}}B_{\Si}^*h_uh_u\!-\!
M_{\bar{\Si}}^2\hs{0.3mm}( \tl l^{\hs{0.4mm}\dag}\!A_{\Si}^*\hs{0.3mm}\tl l^{\hs{0.4mm}*})\!\r ,~~
\bar{\Si}^{\hs{0.4mm}*}\!=\!\frac{1}{M^{2}_{\Si}M_{\bar{\Si}}^2\!-\!|B_{\Si}|^2}
 \!\l \!M^{2}_{\Si}A_{\bar{\Si}}h_uh_u\!+\!
B_{\Si}\hs{0.3mm}(\tl l^{\hs{0.4mm}\dag}\!A_{\Si}^*\hs{0.3mm}\tl l^{\hs{0.4mm}*})\r \!.
\la{sol-Si-barSi}
\eeq
Plugging these solutions back into the potential  (\ref{soft-Si}), we get $\lam_4$-type [of Eq. (\ref{d4})] $d=4$, $\De L=2$ interaction term
\begin{equation}
{\cal L}_{\rm eff}^{\De L=2}=
-\frac{A_{\bar{\Si}}B_{\Si}^*}{M^{2}_{\Si}M_{\bar{\Si}}^2-|B_{\Si}|^2}~(\tl l^{\hs{0.3mm}T}\!A_{\Si}\hs{0.3mm}\tl l) h_uh_u+{\rm h.c}
\la{exact-type2-d4}
\end{equation}
and also quartic term with respect to $\tl l$:
\begin{equation}
{\cal L}^{4\tl l}_{\rm eff}=
\frac{M_{\bar{\Si}}^2}{M^{2}_{\Si}M_{\bar{\Si}}^2-|B_{\Si}|^2}~(\tl l^{\hs{0.3mm}T}\!A_{\Si}\hs{0.3mm}\tl l)
(\tl l^{\hs{0.4mm}\dag}\!A_{\Si}^*\hs{0.3mm}\tl l^{\hs{0.4mm}*})~.
\la{exact-type2-4l}
\eeq
With scales $M_{\Si}M_{\bar{\Si}}\stackrel{>}{_\sim }|B_{\Si}|/3$, the couplings of (\ref{exact-type2-d4})  and (\ref{exact-type2-4l})
  reduce to those given in Eqs. (\ref{lam4-Si}) and (\ref{4slept-ASi}) respectively.

\renewcommand{\theequation}{C.\arabic{equation}}\setcounter{equation}{0}

\section{Evaluating 2-loop Diagram of Fig. \ref{fig-2loop2}-($b$) }
\la{2-loop-calc}
The amplitude, corresponding to the diagram of Fig. \ref{fig-2loop2}-($b$), is given by
{\scriptsize
\beq
{\cal M}=2\cdot 2 \frac{g^{2}g'^{2}}{4} A_{\Sigma}^{ij} B_{\Sigma}^{*} \lambda_{\bar{\Sigma}}
 \int \!\!\frac{%
d^{4}p}{\left( 2\pi \right) ^{4}}\frac{d^{4}q}{\left( 2\pi \right) ^{4}}%
\frac{P_{L}\left( \slashed{p}+M_{\tilde{V}}\right) {\bf C}P_{R}\left( \slashed{p}+\mu\right) P_{R}{\bf C}\left(
\slashed{q}+\mu\right) P_{R}\left( \slashed{q}+M_{\tilde{V}'}\right) {\bf C}P_{L}}{\left(
p^{2}-M_{\tilde{V}}^{2}\right)\!\! \left( p^{2}-\mu^{2}\right) \!\left(
q^{2}-\mu^{2}\right) \!\left( q^{2}-M_{\tilde{V}'}^{2}\right)\! \left(
q^{2}-m_{i}^{2}\right)\! \left( p^{2}-m_{j}^{2}\right) }\frac{1}{\left[ \left(
p+q\right) ^{2}-M_{\Sigma}^{2}\right] \!\left[ \left( p+q\right) ^{2}-M_{\bar{\Sigma}}^{2}\right] }
\la{2-loop-M}
\eeq
}
where ${\bf C}$ and $P_{L,R}$ are charge conjugation and projection matrices respectively.
Using the properties
\beq
\slashed{p}P_{L,R}=P_{R,L}\slashed{p} ~,~~
{\bf C} P_{L,R}=P_{L,R}{\bf C} ~,~~
{\bf C}^{-1}=-{\bf C} ~,~~
P_{R,L}^{2}=P_{R,L} ~,~~
P_{R}P_{L}=0=P_{L}P_{R}
\la{C-PLR-properties}
\eeq
the numerator of the integral in (\ref{2-loop-M}) gets simplified to $\mu^{2}\!\left( pq\right) \!P_{L}{\bf C}^{-1}$ and we obtain:
$$
{\cal M}=\mu^{2} g^{2}g'^{2} A_{\Sigma}^{ij} B_{\Sigma}^{*} \lambda_{\bar{\Sigma}}%
P_{L}{\bf C}^{-1}J~,~~~~~~~{\rm with}
$$
{\footnotesize
\beq
J\!\!=\!\!\int \!\!\frac{d^{4}p}{\left( 2\pi \right) ^{4}}\frac{d^{4}q}{\left( 2\pi
\right) ^{4}}\frac{\left( pq\right) }{\left( p^{2}-M_{\tilde{V}}^{2}\right) \!\!\left(
p^{2}-\mu^{2}\right) \left( q^{2}-\mu^{2}\right) \!\left(
q^{2}-M_{\tilde{V}'}^{2}\right) \!\left( q^{2}-m_{i}^{2}\right) \!\left(
p^{2}-m_{j}^{2}\right) }\frac{1}{\left[ \left( p+q\right) ^{2}-M_{\Sigma
}^{2}\right]\! \left[ \left( p+q\right) ^{2}-M_{\bar{\Sigma}}^{2}\right] }
\la{simpl-M1}
\eeq
}
We can rewrite the fractions of (\ref{simpl-M1}) in the following way
{\footnotesize
$$
\frac{1}{\left( p^{2}-M_{\tilde{V}}^{2}\right)\!\! \left( p^{2}-\mu^{2}\right)
\!\left( p^{2}-m_{j}^{2}\right) }=\!\sum\limits_{n=1}^{3}\frac{A_{n}}{p^2-a_{n}^2}~,~~~~
\frac{1}{\left( q^{2}-M_{\tilde{V}'}^{2}\right) \!\!\left( q^{2}-\mu^{2}\right)
\!\left( q^{2}-m_{i}^{2}\right) }=\!\sum\limits_{m=1}^{3}\frac{B_{m}}{q^2-b_{m}^2} ~,
$$
$$
\frac{1}{\left[ \left( p+q\right) ^{2}-M_{\Sigma }^{2}\right] \left[ \left(
p+q\right) ^{2}-M_{\bar{\Sigma}}^{2}\right] }=\!\! \frac{1}{M_{\Sigma}^{2}-M_{\bar{\Sigma}}^{2}} \sum\limits_{\alpha=1}^{2}\frac{(-1)^{\alpha+1}}{(p+q)^2-c_{\alpha}^2}
$$
}
\beq
{\rm with}~~~~~~~~~~~a_{n}^2=(M_{\tilde{V}}^{2},\mu^{2},m_{j}^{2})~,~~~b_{m}^2=(M_{\tilde{V}'}^{2},\mu^{2},m_{i}^{2}) ~,~~~
c_{\alpha}^{2}=(M_{\Sigma}^{2},M_{\bar{\Sigma}}^{2})
\la{fracs1}
\eeq
and
$$
A_{1}=\frac{1}{( M_{\tilde{V}}^{2}-\mu^{2})( M_{\tilde{V}}^{2}-m_{j}^{2}) }~,~~
A_{2}=\frac{1}{(\mu^{2}-M_{\tilde{V}}^{2})( \mu^{2}-m_{j}^{2}) }~ ,~~
A_{3}=\frac{1}{( m_{j}^{2}-M_{\tilde{V}}^{2})( m_{j}^{2}-\mu^{2}) }~,
$$
\beq
B_{1}=\frac{1}{( M_{\tilde{V}'}^{2}-\mu^{2})(M_{\tilde{V}'}^{2}-m_{i}^{2}) }~, ~~
B_{2}=\frac{1}{(\mu^{2}-M_{\tilde{V}'}^{2})( \mu^{2}-m_{i}^{2}) }~ ,~~
B_{3}=\frac{1}{( m_{i}^{2}-M_{\tilde{V}'}^{2})(m_{i}^{2}-\mu^{2}) } .
\la{AnBm}
\eeq
Note the following properties of the coefficients:
\beq
\sum\limits_{n=1}^{3}A_n = \sum\limits_{m=1}^{3}B_m=0 ~,~~~~~\sum\limits_{n=1}^{3}A_na_n^2 = \sum\limits_{m=1}^{3}B_mb_m^2=0 ~.
\la{sum-prop}
\eeq
Using (\ref{fracs1}), the integral in (\ref{simpl-M1}) can be written as a triple sum:
\beq
 J\!=\!\int \!\!\frac{d^{4}p}{\left( 2\pi \right) ^{4}}\frac{
d^{4}q}{\left( 2\pi \right) ^{4}}\!\!\sum\limits_{\alpha ,n,m}\frac{(-1)^{\alpha+1}A_{n}B_{m}}{M_{\Sigma}^{2}-M_{\bar{\Sigma}}^{2}} \frac{\left( pq\right) }{\left( p^{2}-a_{n}^{2}\right) \left(
q^{2}-b_{m}^{2}\right) \left[ \left( p+q\right) ^{2}-c_{\alpha }^{2}\right] } ~.
\la{simpl-J}
\eeq
Moreover, using the identity
$$
\frac{2\left( pq\right) }{\left( p^{2}-a_{n}^{2}\right) \left(
q^{2}-b_{m}^{2}\right) \![ \left( p+q\right) ^{2}-c_{\alpha }^{2}] }%
=\frac{1}{\left( p^{2}-a_{n}^{2}\right) \left(q^{2}-b_{m}^{2}\right) }-
$$
\beq
-\frac{1}{\left( q^{2}-b_{m}^{2}\right) \![ \left( p+q\right) ^{2}-c_{\alpha }^{2}] }-\frac{1}{%
\left( p^{2}-a_{n}^{2}\right) \![ \left( p+q\right) ^{2}-c_{\alpha }^{2}%
] }+\frac{c_{\alpha }^{2}-a_{n}^{2}-b_{m}^{2}}{\left(
p^{2}-a_{n}^{2}\right) \left( q^{2}-b_{m}^{2}\right) \![ \left(
p+q\right) ^{2}-c_{\alpha }^{2}] } ~,
\la{frac-expand}
\eeq
the identities of Eq. (\ref{sum-prop}) and $\sum\limits_{\al=1}^{2}(-1)^{\al+1}=0$, from (\ref{simpl-J}) we get:
$$
J\!=\!\!\fr{1}{2(M_{\Sigma}^{2}-M_{\bar{\Sigma}}^{2})}\!\sum\limits_{\alpha ,n,m}\!\gamma _{\alpha nm} \!\int \!\frac{d^{4}p}{(2\pi )^{4}}
\frac{d^{4}q}{(2\pi )^4}\frac{1}{(p^{2}-a_{n}^{2})(q^{2}-b_{m}^{2})[ \left( p+q\right) ^{2}-c_{\alpha }^{2}] }
$$
\beq
{\rm with}~~~~~~~\gamma _{\alpha nm}\equiv (-1)^{\alpha+1}(c_{\alpha }^{2}-a_{n}^{2}-b_{m}^{2})A_{n}B_{m} ~.
\la{simpl-J-1}
\eeq
Using the Feynman parametrization
\beq
\frac{1}{d_1d_2d_3}=2\int
\limits_{0}^{1}dx\int\limits_{0}^{1-x}dy\frac{1}{[xd_1+yd_2+(1-x-y)d_3 ]^3}
\la{feyn-par}
\eeq
in the integral of Eq. (\ref{simpl-J-1}),
allows  to perform integration with $p$ and $q$. Due to the properties in (\ref{sum-prop}) and $\sum\limits_{\al=1}^{2}(-1)^{\al+1}=0$,
the divergent parts disappear  (as should be) and we remain with finite result given by:
$$
J=-\frac{1}{2(16\pi^2)^2} \sum\limits_{\alpha ,n,m}\gamma _{\alpha nm} \int\limits_{0}^{1}dx\int\limits_{0}^{1-x}dy
C_{\alpha nm}\ln \left |C_{\alpha nm}\right |
$$
\beq
{\rm with}~~~~~~~~
C_{\alpha nm}=\fr{c_{\al}^2(x+y-1)-a_n^2x-b_m^2y}{(M_{\Sigma}^{2}-M_{\bar{\Sigma}}^{2})(x^2+y^2+xy-x-y)^2} ~.
\la{final-J}
\eeq
Since the amplitude, defined in (\ref{simpl-M1}) (we have calculated), accounts for the operator $\fr{1}{4}(l^T{\cal M}l)h_uh_u$, we
can identify the contribution to the neutrino mass matrix given in Eq. (\ref{2-loop-Mnu}) [for definitions' references see also the comment
after Eq. (\ref{2-loop-Mnu})].

\bibliographystyle{unsrt}

\end{document}